\documentclass{emulateapj}

\usepackage{epsfig,natbib}
\usepackage{graphicx}
\usepackage{amsmath}
\usepackage{booktabs}
\usepackage{longtable}
\usepackage{xspace}
\usepackage{hyperref}
\usepackage[T1]{fontenc}
\usepackage{lipsum}
\newcommand{\mearth}{$M_\earth$~}

\newcommand{\rearth}{$R_\earth$~}

\newcommand{\Searthe}{$S_\earth$}
\newcommand{\rearthe}{$R_\earth$}

\newcommand{\Kepler}{\textit{Kepler}\xspace} 
\newcommand{\Kp}{\textit{Kp}\xspace}

\newcommand{\teff}{\ensuremath{T_{\mathrm{eff}}}\xspace}  
\newcommand{\logg}{\ensuremath{\log g}\xspace} 
\newcommand{\fe}{[Fe/H]\xspace}

\newcommand{\Rp}{\ensuremath{R_P}\xspace}

\newcommand{\Sinc}{\ensuremath{S_{\mathrm{inc}}}\xspace}

\renewcommand{\Re}{\ensuremath{R_{\oplus}}\xspace}

\newcommand{\Rsun}{\ensuremath{R_{\odot}}\xspace }

\usepackage{xstring} 

\newcommand{\val}[1]{%
  \IfEqCase{#1}{%
{nstars-cks}{ 1305 }
{ncand-cks}{2025}
{med-srad-frac-err}{11\%}
{med-smass-frac-err}{ 4\% }
{cks-rp-frac-err-median}{12\%}
{gamma-k}{$k=17.56$}
{gamma-l}{$l=1.00$ (fixed)}
{gamma-theta}{$\theta=0.49$}
{logistic-b}{$b=9.54$}
{logistic-c}{$c=10.09$}
{logistic-d}{$d=0.90$}
{num-sim-planets}{45000}
{num-planets-filtered}{900}
{num-stars-inj}{3840}
{error-bin-fudge}{1.5--2.9}
{nstars-occ}{36,075}
{ncand-sample}{900}
{occ-ratio}{$0.8 \pm 0.2$}
{kstest}{0.003}
{adtest}{0.012}
{tr-prob-sm}{6\%}
{tr-prob-lg}{3\%}
{det-prob-sm}{86\%}
{det-prob-lg}{96\%}
}[\PackageError{tree}{Undefined option to tree: #1}{}]%
}

\shortauthors{Fulton {et~al.}}
\shorttitle{The Radius Gap}
\submitted{Accepted for publication in the Astronomical Journal}
\begin{document}
\pagenumbering{arabic}


\title{The California-Kepler Survey. \\ 
III.  A Gap in the Radius Distribution of Small Planets\altaffilmark{1}}
\author{
Benjamin J.\ Fulton\altaffilmark{2,3,12,*},
Erik A.\ Petigura\altaffilmark{3,15},
Andrew W.\ Howard\altaffilmark{3},
Howard Isaacson\altaffilmark{4},
Geoffrey W. Marcy\altaffilmark{4},
Phillip A.\ Cargile\altaffilmark{5},
Leslie Hebb\altaffilmark{6}, 
Lauren M.\ Weiss\altaffilmark{7,13},
John Asher Johnson\altaffilmark{5},
Timothy D.\ Morton\altaffilmark{8},
Evan Sinukoff\altaffilmark{2,3,14},
Ian J.\ M.\ Crossfield\altaffilmark{9,16},
Lea A.\ Hirsch\altaffilmark{3}
}

\altaffiltext{1}{Based on observations obtained at the W.\,M.\,Keck Observatory, 
                      which is operated jointly by the University of California and the 
                      California Institute of Technology.  
                      Keck time was granted for this project by 
                    the University of California, and California Institute of Technology, the University of Hawaii, and NASA.
                      }  
\altaffiltext{2}{Institute for Astronomy, University of Hawai`i, 2680 Woodlawn Drive, Honolulu, HI 96822, USA} 
\altaffiltext{3}{California Institute of Technology, Pasadena, California, U.S.A.}
\altaffiltext{4}{Department of Astronomy, University of California, Berkeley, CA 94720, USA}
\altaffiltext{5}{Harvard-Smithsonian Center for Astrophysics, 60 Garden St, Cambridge, MA 02138, USA}
\altaffiltext{6}{Hobart and William Smith Colleges, Geneva, NY 14456, USA}
\altaffiltext{7}{Institut de Recherche sur les Exoplan\`{e}tes, Universit\'{e} de Montr\'{e}al, Montr\'{e}al, QC, Canada}
\altaffiltext{8}{Department of Astrophysical Sciences, Peyton Hall, 4 Ivy Lane, Princeton, NJ 08540 USA}
\altaffiltext{9}{Astronomy and Astrophysics Department, University of California, Santa Cruz, CA, USA}

\altaffiltext{12}{National Science Foundation Graduate Research Fellow}
\altaffiltext{13}{Trottier Fellow}
\altaffiltext{14}{Natural Sciences and Engineering Research Council of Canada Graduate Student Fellow}
\altaffiltext{15}{Hubble Fellow}
\altaffiltext{16}{NASA Sagan Fellow}
\altaffiltext{*}{bfulton@hawaii.edu}

\begin{abstract}
The size of a planet is an observable property directly connected to the physics of its formation and evolution.  
We used precise radius measurements from the California-Kepler Survey (CKS) to study the size distribution of \val{ncand-cks} \Kepler planets in fine detail.  
We detect a factor of $\geq$2 deficit in the occurrence rate distribution at 1.5--2.0 \rearthe. This gap splits the population of close-in ($P$ < 100 d) small planets into two size regimes: $R_P < $~1.5 \rearth and $R_P = 2.0$--$3.0$~\rearthe, with few planets in between.  
Planets in these two regimes have nearly the same intrinsic frequency based on occurrence measurements that account for planet detection efficiencies.
The paucity of planets between 1.5 and 2.0 \rearth supports the emerging picture that close-in planets smaller than Neptune are composed of rocky cores measuring 1.5 \rearth or smaller with varying amounts of low-density gas that determine their total sizes. 
\end{abstract}

\section{Introduction}
\label{sec:intro}

NASA's \Kepler space telescope enabled the discovery of over 4000 transiting planet candidates\footnote{NASA Exoplanet Archive, 2/27/2017}$^,$\footnote{
The false positive probability for the majority of the \Kepler candidates is 5--10\% \citep{Morton11}.} opened the door to detailed studies of exoplanet demographics. One of the first surprises to arise from studies of the newly revealed sample of planets was the multitude of planets with radii smaller than Neptune but larger than Earth \citep[$R_P$=1.0--3.9 \rearthe,][]{Batalha13}. Our solar system has no example of these intermediate planets, yet they are by far the most common in the \Kepler sample \citep{Howard12,Fressin13,Petigura13a, Youdin11, Christiansen15, Dressing15, Morton14}.

A key early question of the \Kepler mission was whether these sub-Neptune-size planets are predominantly rocky or possess low-density envelopes that contribute significantly to the planet's overall size. The radial velocity (RV) follow-up effort of the \Kepler project focused on 22 stars hosting one or more sub-Neptunes \citep{Marcy14}. In addition, detailed modeling of transit timing variations (TTVs) provided mass constraints for a large number of systems in specific architectures \citep[e.g.,][]{Wu13, Hadden14, Hadden16}. The resulting mass measurements revealed that most planets larger than 1.6 \rearth have low densities that were inconsistent with purely rocky compositions, and instead required gaseous envelopes \citep{Weiss14,Rogers15}.

The distinction between rocky and gaseous planets reflects the typical core sizes of planets as well as the physical mechanisms by which planets acquire (and lose) gaseous envelopes. The densities of planets with radii smaller than $\sim$1.6 \rearth are generally consistent with a purely rocky composition \citep{Weiss14, Rogers15} and their radius distribution likely reflects their initial core sizes. However, a small amount of H/He gas added to a roughly Earth-size rocky core can substantially increase planet size, without significantly increasing planet mass. For this reason, it has been suggested that the radii of sub-Neptune-size planets, along with knowledge of the irradiation history, would be sufficient to estimate bulk composition without additional information \citep{Lopez13b, Wolfgang15}.

The large number of planets smaller than Neptune discovered by the \Kepler mission was unexpected given prevailing theories of planet formation, which were developed to explain the distribution of giant planets \citep{Ida04, Mordasini09}. These theories predicted that planets should either fail to accrete enough material to become super-Earths, or they would grow quickly, accreting all of the gas in their feeding zones growing to massive, gas-rich giant planets. Modern formation models are now able to reproduce the observed population of super-Earths \citep{Hansen12, Mordasini12, Alibert13, Chiang13, Lee14, Chatterjee14, Coleman14, Raymond14, Lee16}. Many of these new models can be corroborated by measuring the bulk properties of individual planets and the typical properties of the population.

As formation models continue to be refined, the role of atmospheric erosion on these short-period planets is becoming more apparent. Several authors have predicted the existence of a ``photoevaporation valley'' in the distribution of planet radii \citep[e.g.,][]{Owen13,Lopez13,Jin14,Chen16,Lopez16}.

Photoevaporation models predict that there should be a dearth of intermediate sub-Neptune size planets orbiting in highly irradiated environments. The mass of H/He in the envelope must be finely tuned to produce a planet in this intermediate size range. Planets with too little gas in their envelopes are stripped to bare, rocky cores by the radiation from their host stars. In general, the radii of bare, rocky cores versus planets with a few percent by mass H/He envelopes depend on many uncertain variables such as the initial core mass distribution and the insolation flux received by the planet. A rift in the distribution of small planet radii is a common result of the planet formation models that include photoevaporation.

\citet{Owen13} provided tentative observational evidence for such a feature in the radius distribution of \Kepler planets. They observed a bimodal structure in the planet radius distribution, particularly when the planet sample was split into subsamples with low and high integrated X-ray exposure histories.  However, the relatively large planet radius uncertainties in \citet{Owen13} diluted the gap and reduced its statistical significance.  Their study also considered the number distribution  of planets, and was not corrected for completeness as we do below.  Such corrections mitigate sample bias and allow for the recovery of the underlying planet distribution from the observed one.

Here, we examine a sample of planets orbiting stars with precisely measured radii from the California-Kepler Survey (CKS; see \citet{Petigura17} and \citet{Johnson17}). We use the precise stellar radii to update the planet radii, bringing the distribution of planet radii into sharper focus and revealing a gap between 1.5 and 2.0 \rearthe. 

This paper is structured as follows. In \S \ref{sec:data} we discuss our stellar and planetary samples. We describe our methods for correcting for pipeline search sensitivity and transit probabilities in \S \ref{sec:completeness}. In \S \ref{sec:valley} we examine the one-dimensional marginalized radius distribution and also two-dimensional distributions of planet radius as a function of orbital period, stellar radius, and insolation flux.  We discuss potential explanations for the observed planet radius gap in \S \ref{sec:discussion} and finish with some concluding remarks in \S \ref{sec:conclusion}.

\section{Sample of Planets}
\label{sec:data}
\subsection{California \Kepler Survey}

For this work we adopt the stellar sample and the measured stellar parameters from the CKS program \citep[][hereafter Paper I]{Petigura17}. The measured values of \teff, \logg, and \fe are based on a detailed spectroscopic characterization of \val{nstars-cks} \Kepler Object of Interest (KOI) host stars using observations from Keck/HIRES \citep{Vogt94}. In \citet[][hereafter Paper II]{Johnson17}, we associated those stellar parameters from Paper I to Dartmouth isochrones \citep{Dotter08} to derive improved stellar radii and masses, allowing us to recalculate planetary radii using the light curve parameters from \cite{Mullally15}, hereafter ``Q16''. Median uncertainties in stellar radius improve from 25\% \citep[][]{Huber14} to \val{med-srad-frac-err} after our CKS spectroscopic analysis. Stellar mass uncertainties improve from 14\% to \val{med-smass-frac-err} in the Paper II catalog. This leads to median uncertainties in planet radii  of \val{cks-rp-frac-err-median} which enable the detection of finer structures in the planet radius distribution.

\subsection{Sample Selection}
\label{sec:sample}

The CKS stellar sample was constructed to address a variety of science topics (Paper I). The core sample is a magnitude-limited set of KOIs (\Kp $<$ 14.2). Additional fainter stars were added to include habitable zone planets, ultra-short-period planets, and multi-planet systems. Here, we enumerate a list of cuts in parameter space designed to create a sample of planets with well-measured radii and with well-quantified detection completeness. The primary goal is to determine anew the occurrence of planets as a function of planet radius, with greater reliability than was previously possible.

We start by removing planet candidates deemed false positives in Paper I. The Paper I false positive designations were determined using the false positive probabilities calculated by \citet{Morton11, Morton12, Morton16}, the \Kepler team's designation available on the NASA Exoplanet Archive, and a search for secondary lines in the HIRES spectra \citep{Kolbl15} as well as any other information available in the literature for individual KOIs. Next, we restrict our sample to only the magnitude-limited portion of the larger CKS sample (\Kp $ < 14.2$).

The planet-to-star radius ratio ($R_P/R_{\star}$) becomes uncertain at high impact parameters ($b$) due to degeneracies with limb-darkening. We excluded KOIs with $b > 0.7$ to minimize the impact of grazing geometries. We experimented other thresholds in $b$ and found that our results are relatively insensititve to $b <$ 0.6, 0.7, or 0.8, with the trade-off of smaller sample size with decreasing threshold in $b$.

We removed planets with orbital periods longer than 100 days in order to avoid domains of low completeness (especially for planets smaller than about 4 \rearthe) and low transit probability.

\begin{figure}[ht]
\includegraphics[scale=0.3]{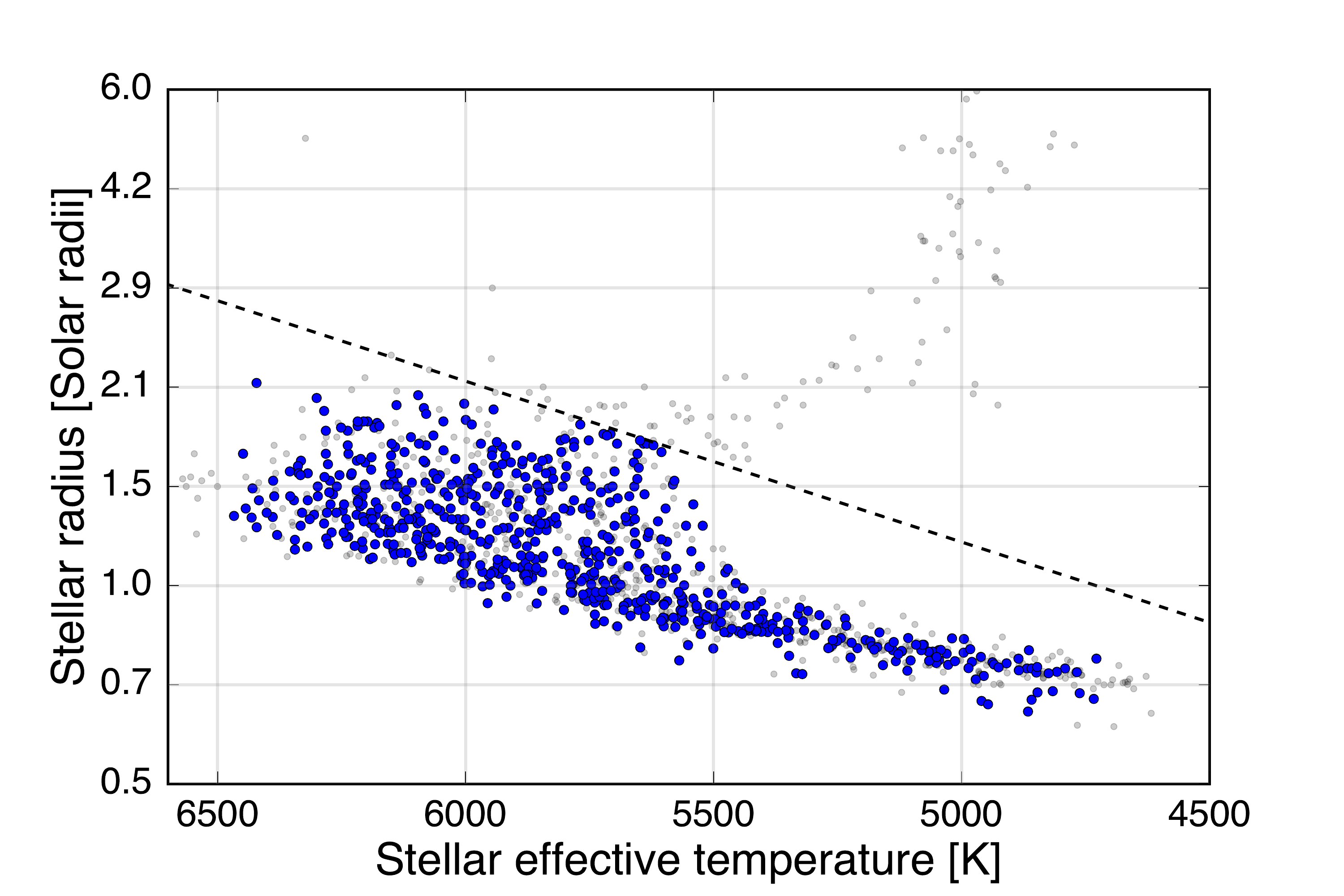}
\includegraphics[scale=0.3]{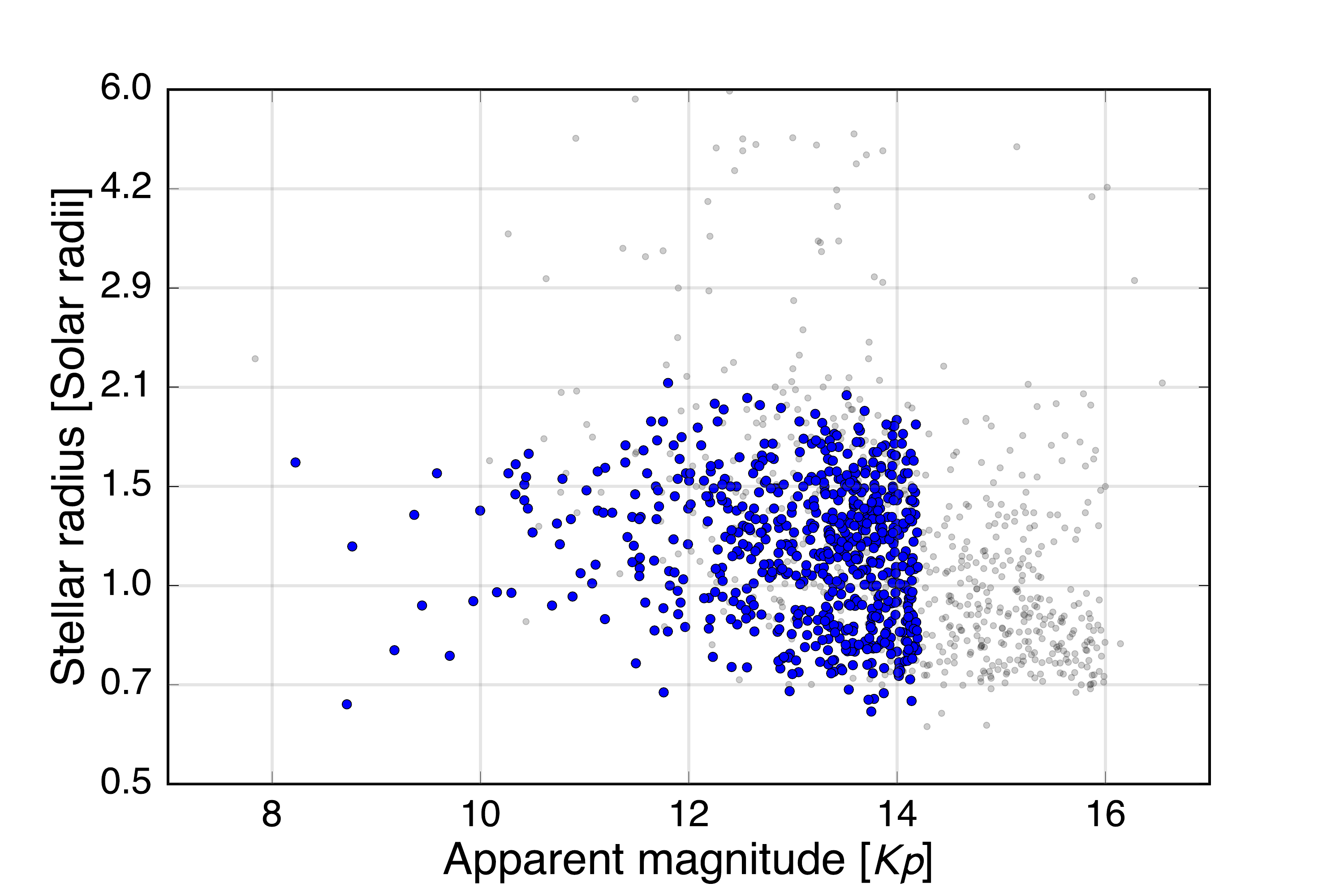}
\caption{\emph{Top:} HR diagram of the sample of stars selected for analysis. The full Paper II sample is plotted in light grey points and the sample selected for analysis after applying the filters discussed in Section \ref{sec:sample} are plotted as blue squares. Giant planet hosting stars that fall above the dashed line given by Equation \ref{eqn:giants} are omitted from the final sample.
\emph{Bottom:} Stellar radius of CKS stars as a function of \Kepler magnitude (\Kp). We note that stars fainter than 14.2 do not follow the same stellar radius distribution. We omit stars fainter than \Kp$= 14.2$ to avoid biasing our planet radius distribution. The point colors are the same as in the \emph{top} panel.
}
\label{fig:HR}
\end{figure}

We also excised planets orbiting evolved stars since they have somewhat lower detectability and less certain radii.  This was implemented using an \emph{ad hoc} temperature-dependent stellar radius filter,  
\begin{equation}
\frac{R_{\star}}{R_{\odot}} > 10^{0.00025(T_\mathrm{eff}/{\rm K} - 5500) + 0.20},
\label{eqn:giants}
\end{equation}
which is plotted in Figure \ref{fig:HR}.
We also restricted our sample to planets orbiting stars within the temperature range where we can extract precise stellar parameters from our high resolution optical spectra (6500--4700 K).
Finally, we accounted for uncertainties in the completeness corrections caused by systematic and random measurement errors in the simulations, described in Appendix \ref{sec:validation}.

\begin{deluxetable}{lr}
\tabletypesize{\small}
\tablecaption{Depth of the Gap}
\tablehead{ 
    \colhead{Filter}   &   \colhead{$V_A$} \\
    \colhead{}   &    \colhead{}
}
\startdata
Full CKS sample  &  0.746  \\
False positives removed  &  0.742  \\
\Kp$ < 14.2$  &  0.686  \\
$b < 0.7$  &  0.572  \\
$P < 100$ d  &  0.498  \\
Giant stars removed  &  0.507  \\
$T_{\rm eff}$ = 4700--6500 K  &  0.483 \\
\enddata
\label{tab:va}
\end{deluxetable}

\begin{figure}
\includegraphics[scale=0.38]{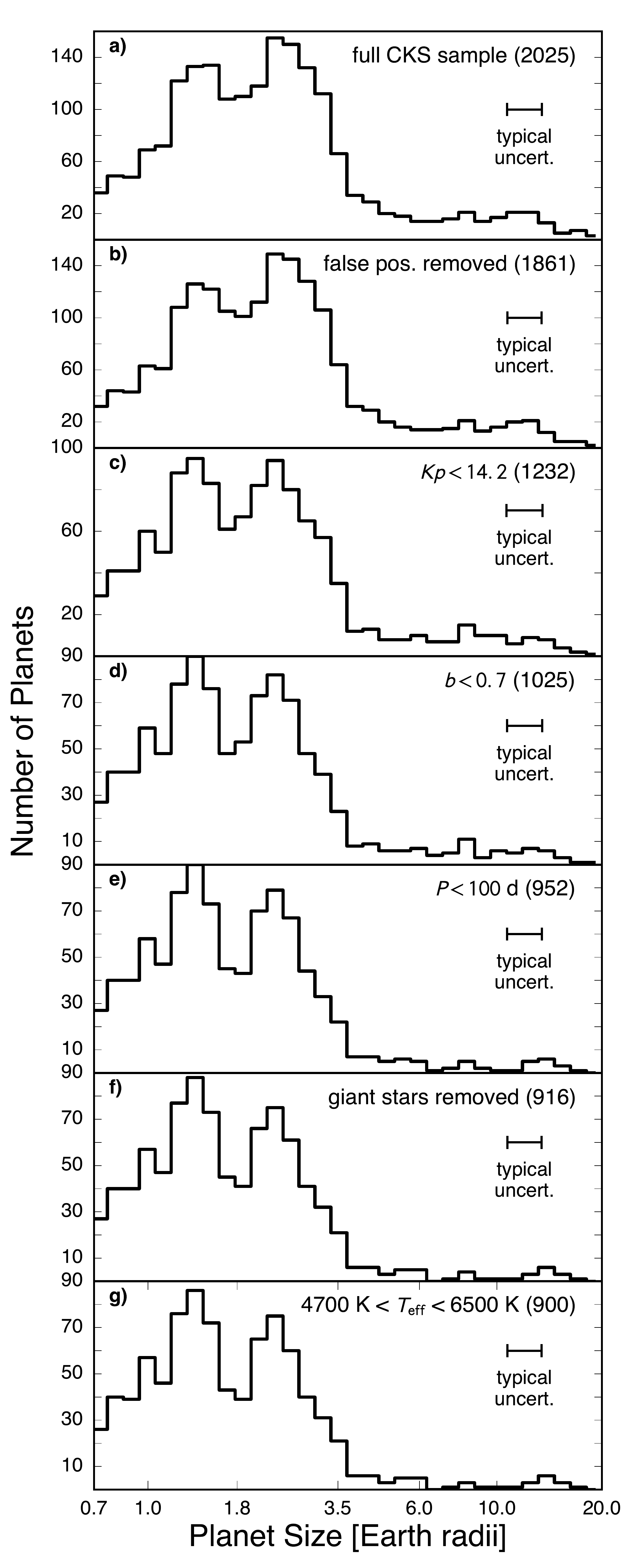}
\caption{
{\bf(a)} Size distribution of all planet candidates in the CKS planet sample.
Panels {\bf(b)--(g)} show the radius distribution after applying several successive cuts to {\bf(b)}: remove known false positives, {\bf(c)}: keep candidates orbiting bright stars (\Kp < 14.2), {\bf(d)}: retain candidates with low impact parameters ($b < 0.7$), {\bf(e)}: keep candidates with orbital periods shorter than 100 days, {\bf(f)}: remove candidates orbiting giant host stars, and {\bf(g)}: include only candidates orbiting stars within our adopted $T_{\rm eff}$ range (4700 K $<$ $T_{\rm eff}$ $<$ 6500 K).
The number of planets remaining after applying each successive filter is annotated in the upper right portion of each panel.
Our filters produce a reliable sample of accurate planet radii and accentuate the deficit of planets at 1.8 \rearthe.
}
\label{fig:filters}
\end{figure}

The multiple filters purify the CKS sample of stars and planets and are summarized in Figure \ref{fig:filters}. We assessed the impact of filters on the depth of the planet radius valley using an \textit{ad hoc} metric $V_A$. This quantity is defined as the ratio of the number of planets with radii of 1.64--1.97 \rearth (the bottom of the valley) to the average number of planets with radii of 1.2--1.44 \rearth or 2.16--2.62 \rearthe (the peaks of the distrubtion immediately outside of the valley).
The radius limits for the calculation of $V_A$ were chosen so that $V_A=1$ for a log-uniform distribution of planets with radii between 1.2 \rearth and 2.62 \rearthe. Smaller values of $V_A$ denote a deeper valley.  
The values of $V_A$ after applying each successive filter are tabulated in Table \ref{tab:va}.

\citet{Furlan17} compiled a catalog of KOI host stars that were observed using a collection of high-resolution imaging facilities \citep{Lillo-Box12, Lillo-Box14, Horch12, Horch14, Everett15, Gilliland15, Cartier15, Wang15a, Wang15b, Adams12, Adams13, Dressing14, Law14, Baranec16, Howell11}. Many of the 1902 KOIs in the \citet{Furlan17} catalog also appear in our sample. We investigated removing KOI hosts with known companions or large dilution corrections but found no significant changes to the shape of the distribution. Since only a subset of our KOIs were observed by \citet{Furlan17} and it is difficult to determine the binarity of the parent stellar population for occurrence calculations, we chose not to filter our planet catalog using the results of high-resolution imaging. However, many of these stars may have already been identified as false positives in the Paper I catalog and therefore removed from our final sample of planets.

\begin{figure}
\includegraphics[scale=0.37]{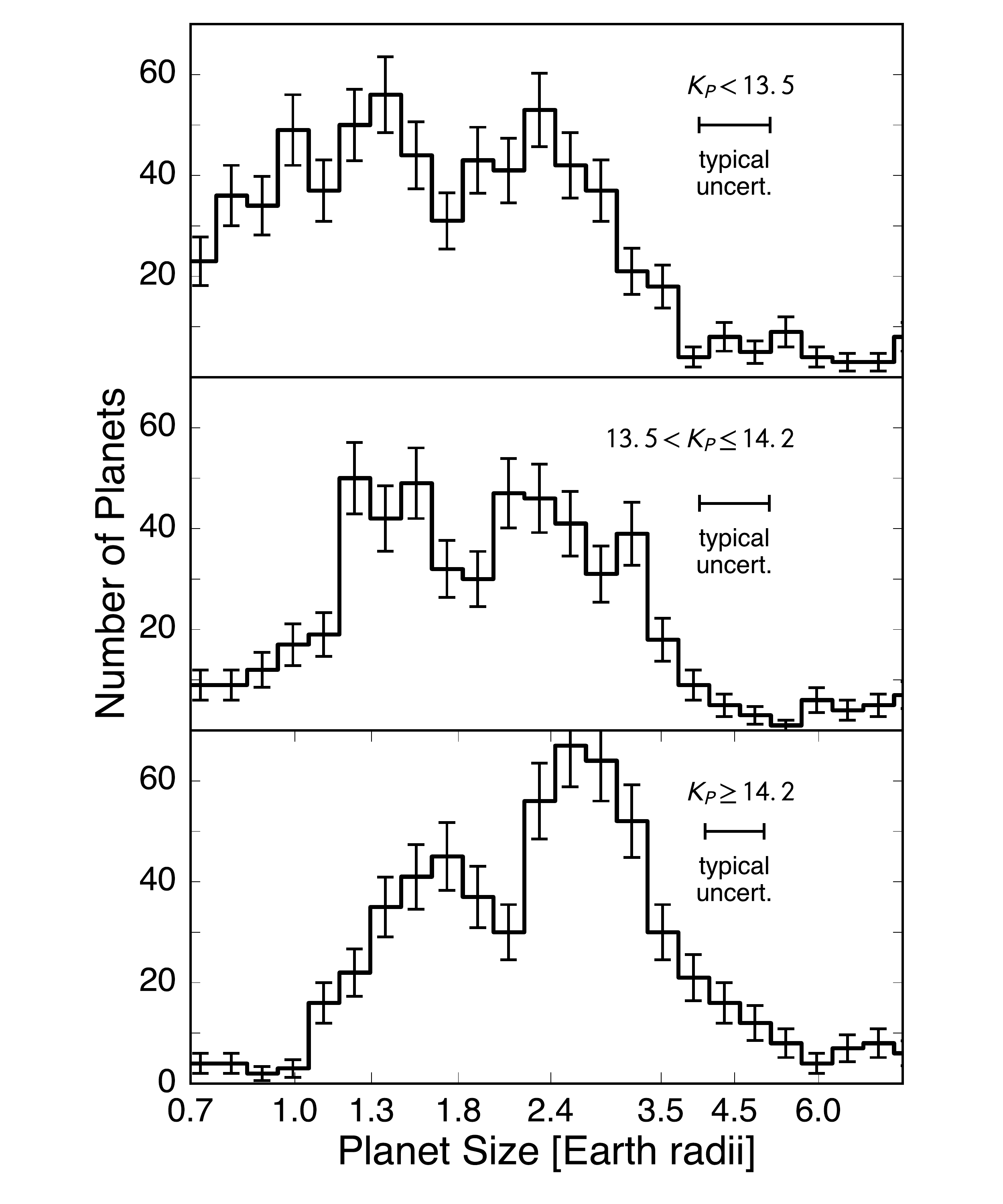}
\caption{Histograms of planet radii broken up into the three magnitude ranges annotated in each panel. All of the filters have been applied to the sample as described in \S \ref{sec:sample}.  
The gap is apparent in all magnitude ranges. The distribution of planet radii in the two brightest magnitude ranges are indistinguishable (p-value = 0.6).
However, the planets orbiting stars with \Kp $>$ 14.2 are statistically different (p-value = 0.0004) when compared to the \Kp = 13.5--14.2 magnitude range. This is expected due to the non-systematic nature of the target selection for CKS and KIC stars fainter than \Kp = 14.2. This motivates our removal of planets with hosts fainter than \Kp = 14.2.
}
\label{fig:maghists}
\end{figure}

We investigated the impact of our apparent magnitude cut by examinging the size distribution for three ranges of \Kp  (Figure \ref{fig:maghists}). For these tests we applied all of the filters described in this section except the \Kp$ < 14.2$ magnitude cut. We found that the planet radius distribution for \Kp$ < 13.5$ is statistically indistinguishable from the radius distribution for planets orbiting stars with $13.5 <$ \Kp $\leq 14.2$. An Anderson-Darling test \citep{Anderson52,Scholz87} predicts that the two distributions were drawn from the same parent population with a p-value of 0.6. However, the radius distribution of planets orbiting host stars with $\Kp \geq 14.2$ is visually and statistically different (p-value < 0.0004). This is somewhat expected given the non-systematic target selection for both the initial \Kepler target stars and the stars observed in the CKS survey. Stars with \Kp$ > 14.2$ were only observed in the CKS program because they were hosts to multi-planet systems, habitable-zone candidates, ultra-short period planets, or other special cases. Targets fainter than $\Kp >14.0$ were observed by \Kepler only if their stellar and noise properties indicated that there was a high probability of the detection of small planets \citep{Batalha10}. These non-uniform \Kepler target selection effects motivate our choice to exclude faint stars. The final distributions of planet radii do not depend on the \Kp$ < 14.2$ or \Kp$ < 14.0$ (p-value $>$ 0.95) choice. But there are 153 planet candidates with $14.0 < $\Kp$ < 14.2$ so we choose to include those additional candidates to maximize the statistical power of the final sample.

\begin{figure}
\centering
\includegraphics[scale=0.31]{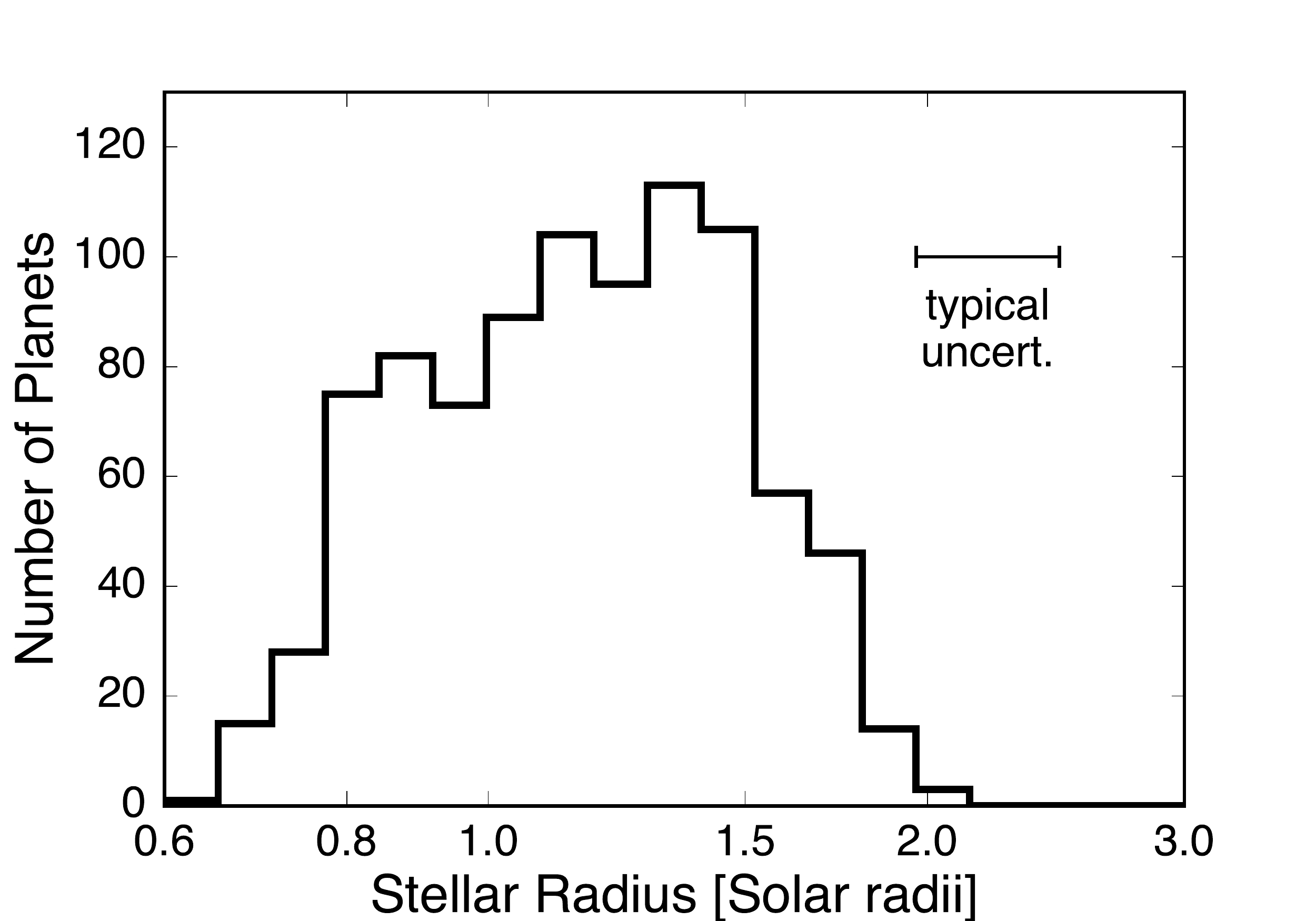}
\includegraphics[scale=0.31]{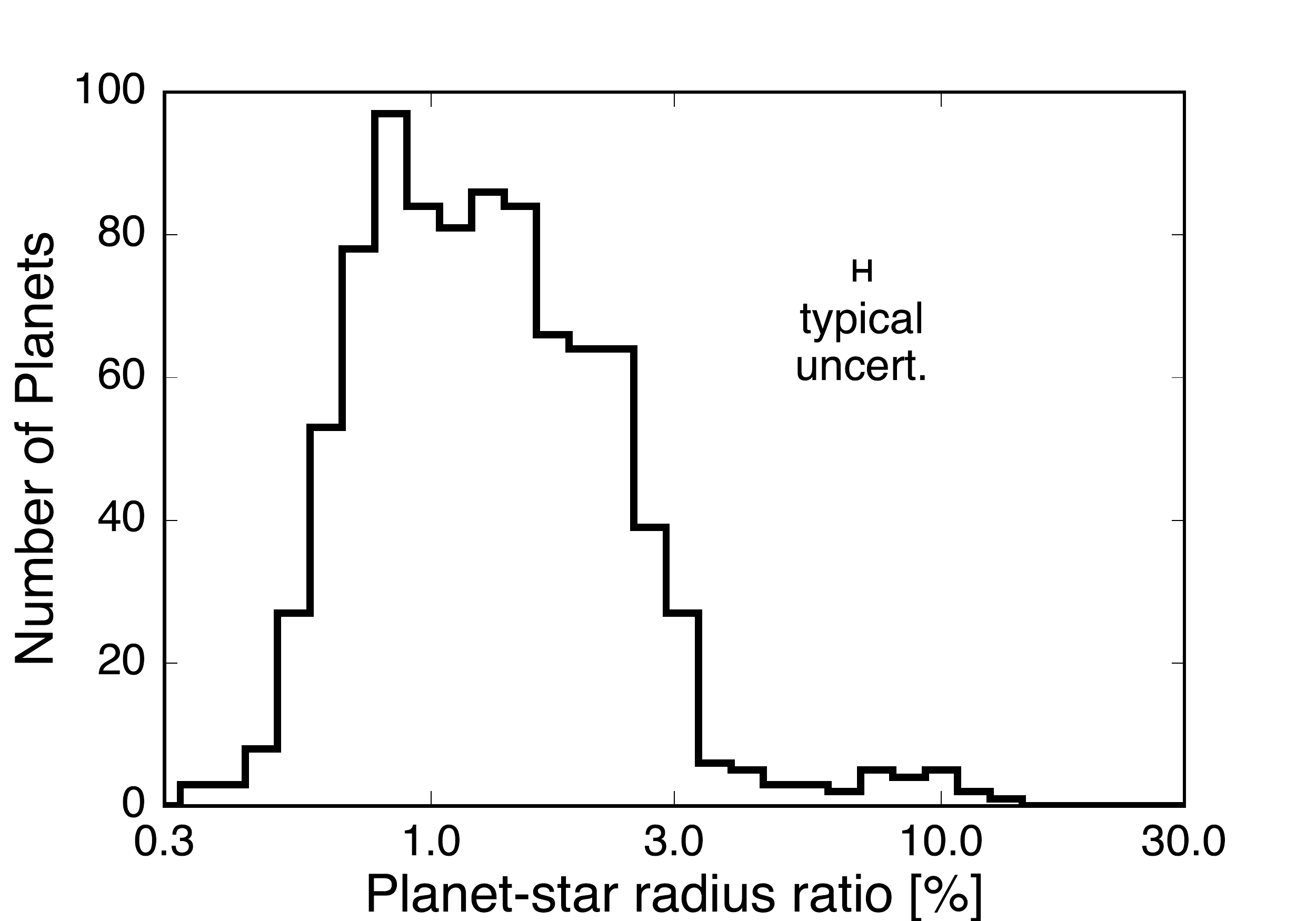}
\caption{\emph{Top:} Histogram of stellar radii derived in Paper II and used to update planet radii in this work after the filters described in Section \ref{sec:sample} are applied.
\emph{Bottom:} Histogram of planet-to-star radii ratios for the stars remaining after the filters described in Section \ref{sec:sample} are applied to the full Paper II sample of planet candidates. In both cases, the median measurement uncertainties are plotted in the upper right. Neither of these two histograms shows the same bimodal feature that is observed in the planet radius distribution, which demonstrates that the feature is not an artifact of our stellar sample or transit fitting.}
\label{fig:rprs}
\end{figure}

The two distinct peaks separated by a valley (Figure \ref{fig:filters}) are apparent in the initial number distribution of planet radii and the final distribution after the filters are applied. The depth of the valley increases as we apply these filters, suggesting that the purity of the planet sample improves with filter application.  Note that the filters act on the stellar characteristics and are agnostic to planet radius.

Figure \ref{fig:rprs} shows histograms of the stellar radii and planet-to-star radius ratios ($R_{P}/R_{\star}$) for the filtered sample stars. These two distributions are both unimodel. This demonstrates that the bimodality of the planet radius distribution is not an artifact of the stellar sample or the light curve fitting used to measure $R_{P}/R_{\star}$.

\section{Completeness Corrections}
\label{sec:completeness}

To recover the underlying planet radius distribution from the observed distribution we made completeness corrections to compensate for decreasing detectability of planets with small radii and/or long orbital periods.

An additional complication associated with the completeness corrections in this work is that the stellar properties of the planet-hosting stars come from a different source and have higher precision than the stellar properties for the full set of \Kepler\ target stars. We explore the additional uncertainties introduced by this fact by running a suite of simulated transit surveys described in Appendix \ref{sec:validation}. We inflate the uncertainties on the histogram bin heights  by the scaling factors listed in Table \ref{tab:unc} to account for these effects.

\subsection{Pipeline Efficiency}

We followed the procedure described in \citet{Christiansen16} using the results from their injection-recovery experiments \citep{Christiansen15}. They injected about ten-thousand transit signals into the raw pixel data and processed the results with version 9.1 of the official \Kepler pipeline \citep{Jenkins13}. These completeness tests were used to identify combinations of transit light curve parameters that could be recovered by the \Kepler pipeline for a given sample of target stars. They injected signals onto both target stars and neighboring pixels to quantify the pipeline's ability to identify astrophysical false positives. We assumed that our sample is free of the vast majority of false positives so we only considered injections of transits onto the target stars. We only considered injections on stars that would have been included in the CKS sample and would not be removed by the filters described in \S \ref{sec:sample}. Namely, we considered injected impact parameters less than 0.7, injected periods shorter than 100 days, \Kp$ \leq 14.2$, 4700 K $<\teff<$ 6500 K, and stellar radii compatible with Equation \ref{eqn:giants} based on the values in the Stellar17 catalog\footnote{https://archive.stsci.edu/kepler/stellar17/search.php} prepared by the \Kepler stellar parameters working group \citep{Mathur16}.
This leaves a total of \val{num-stars-inj} synthetic transit signals injected onto the target pixels of \val{num-stars-inj} stars observed by \Kepler.
We also apply these same filters to the stars in the Stellar17 catalog. The number of stars remaining after the filters are applied is the number of stars observed by \Kepler that could have led to detections of planets that would be present in our filtered planet catalog ($N_{\star}=$~\val{nstars-occ}).
We calculated the fraction of injected signals recovered as a function of injected signal-to-noise as
\begin{equation}
m_i = \left(\frac{R_P}{R_{\star,i}}\right)^2 \sqrt{\frac{T_{\rm obs, \emph{i}}}{P}} \left(\frac{1}{\rm{CDPP}_{\rm dur, \emph{i}}}\right),
\label{eqn:snr}
\end{equation}
where $R_{P}$ and $P$ are the radius and period of the particular injected planet. $R_{\star,i}$ is the stellar radius for the $i^{\rm th}$ star in the Stellar17 catalog,
$T_{\rm obs, \emph{i}}$ is the amount of time that the particular star was observed, and $\rm{CDPP}_{\rm dur, \emph{i}}$ is the Combined Differential Photometric Precision \citep[CDDP,][]{Koch10} value for each star extrapolated to the transit duration for each injection. We fit a 2$^{\rm nd}$ order polynomial in $1/\sqrt{d}$ to the $d=3$, 6, and 12-hour CDPP values for each star to perform the extrapolation \citep{Sinukoff13}.

\begin{figure}
\centering
\includegraphics[scale=0.3]{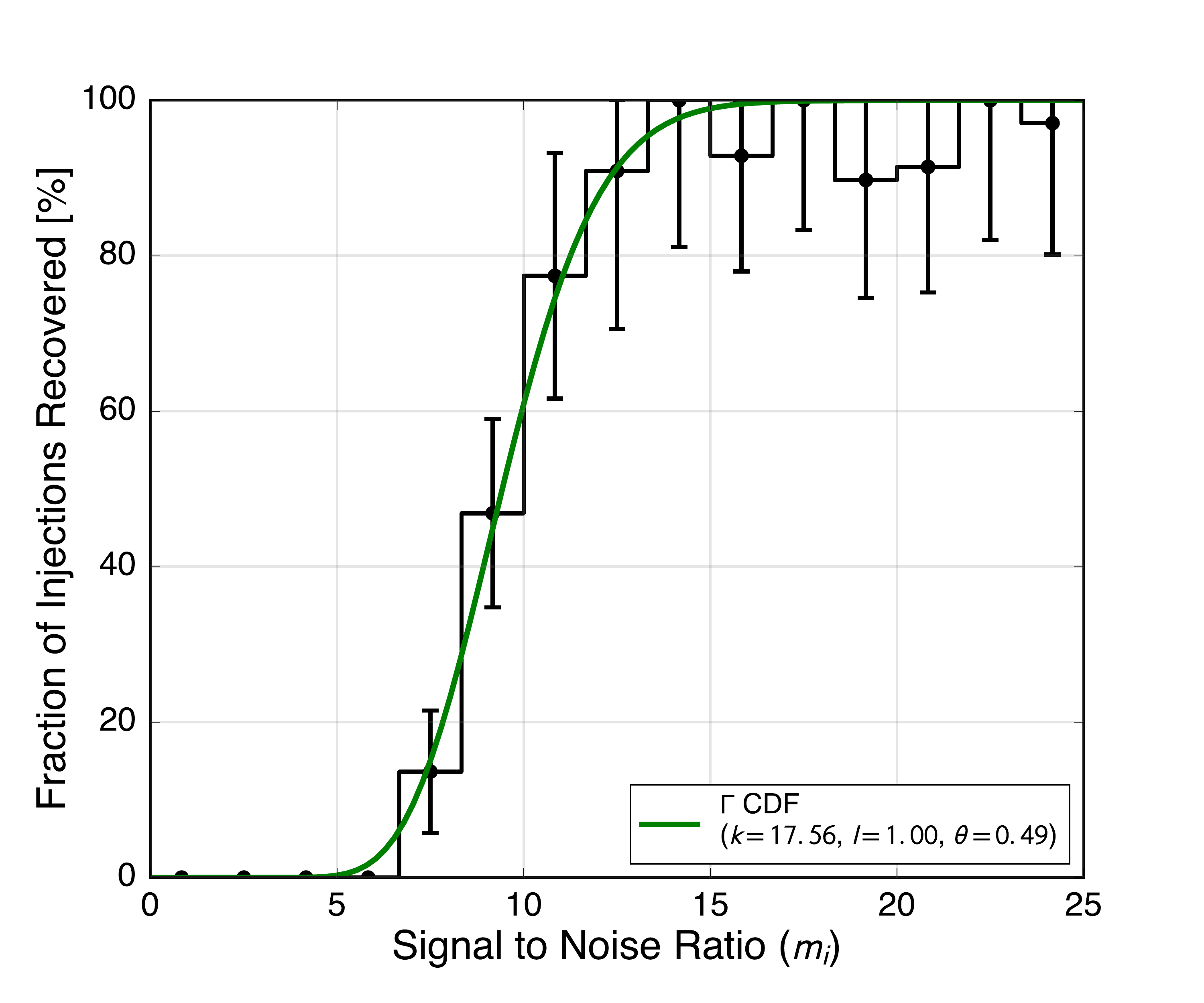}
\caption{Fraction of injected transit signals recovered as a function of signal to noise ratio ($m_i$, Equation \ref{eqn:snr}) in our subsample of the \Kepler target stars using the injection recovery tests from \citet{Christiansen15}. We fit a $\Gamma$ CDF (Equation \ref{eqn:gamma}) and plot the best-fit model in green. }
\label{fig:gammacdf}
\end{figure}

We fit a $\Gamma$ cumulative distribution function (CDF) to the recovery fraction vs. injected ($m_i$) of the form
\begin{equation}
C(m_i; k, \theta, l) = \Gamma(k) \int_0^{\frac{m_i-l}{\theta}} t^{k-1} e^{-t} dt,
\label{eqn:gamma}
\end{equation}
to derive the average pipeline efficiency. $C(m_i)$ is the probability that a signal with a given value of $m_i$ would actually be detected by the \Kepler transit search pipeline. In practice we used the \texttt{scipy.stats.gammacdf}(t, k, l, $\theta$) function in SciPy version 0.18.1. Using the \texttt{lmfit} Python package \citep{Newville14} to minimize the residuals we found best-fit values of \val{gamma-k}, \val{gamma-l}, and \val{gamma-theta}. Figure \ref{fig:gammacdf} shows the fraction of injections recovered as a function of $m_i$ and our model for pipeline efficiency.

Our pipeline efficiency curve is $\sim$15-25\% lower than the efficiency as a function of the \Kepler multi-event statistic (MES) derived in \citep{Christiansen15} for their FGK subsample. The difference can be explained by the fact that the MES is estimated in the \Kepler pipeline during a multidimensional grid search. In most cases, the search grid is not fine enough to find the exact period and transit time for a given planet candidate. Since the grid search doesn't find the best-fit transit model it generally underestimates the SNR ($m_i$) by a factor of $\sim$25\% (Petigura et al., in preparation).

\subsection{Survey Sensitivity}
\label{sec:sensitivity}

\begin{deluxetable*}{lrrrrrrr}
\tablecaption{Planet Detection Statistics\label{tab:planet}}
\tabletypesize{\scriptsize}
\tablecolumns{7}
\tablewidth{0pt}
\tablehead{
	\colhead{Planet} & 
	\colhead{$P$} &
	\colhead{\Rp} & 
	\colhead{SNR} &
	\colhead{Detection probability} & 
    \colhead{Transit probability} &
    \colhead{Weight}
    \\
    \colhead{candidate} & 
	\colhead{d} &
	\colhead{\Re} & 
	\colhead{$m_{i}$} &
	\colhead{$p_{\rm det}$} & 
    \colhead{$p_{\rm det}$} &
    \colhead{$1/w_i$}
}
\startdata
 K00002.01 &  2.20 & 13.41 &  750.22 & 1.00 & 0.14 &  6.94 \\
 K00003.01 &  4.89 &  5.11 &  877.10 & 1.00 & 0.05 & 20.14 \\
 K00007.01 &  3.21 &  4.13 &  146.38 & 1.00 & 0.11 &  8.88 \\
 K00010.01 &  3.52 & 13.39 &  914.62 & 1.00 & 0.09 & 11.06 \\
 K00017.01 &  3.23 & 15.04 & 1212.38 & 1.00 & 0.11 &  9.40 \\
 K00018.01 &  3.55 & 13.94 &  820.96 & 1.00 & 0.10 &  9.58 \\
 K00020.01 &  4.44 & 21.41 & 1469.42 & 1.00 & 0.10 & 10.15 \\
 K00022.01 &  7.89 & 14.20 & 1085.97 & 1.00 & 0.06 & 17.98 \\
 K00041.01 & 12.82 &  2.37 &   37.15 & 0.98 & 0.05 & 22.37 \\
 K00041.02 &  6.89 &  1.35 &   15.04 & 0.91 & 0.07 & 15.98 \\

\enddata
\label{tab:weights}
\tablecomments{Table \ref{tab:weights} is available in its entirety in machine-readable format, which also includes period and radius uncertainties.
A portion is shown here for guidance regarding its form and content.
Refer to Paper II for the CKS stellar parameters associated with each KOI. This table contains only the subset of planet detections that passed the filters described in \S \ref{sec:sample}.
The full sample of planet candidates orbiting CKS target stars can be found in Paper II.}
\end{deluxetable*}

\begin{figure}
\centering
\includegraphics[width=1.1\linewidth]{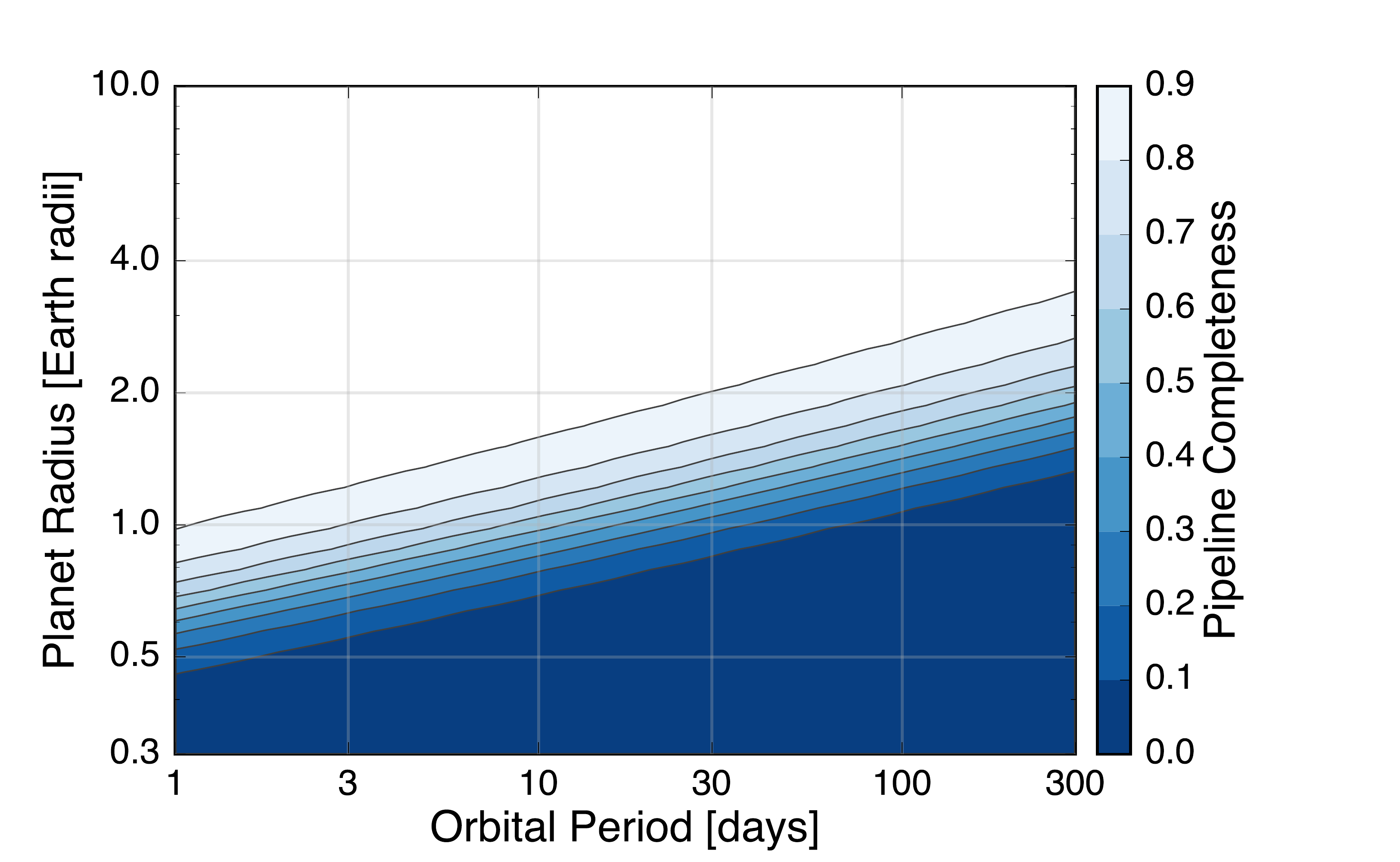}
\includegraphics[width=1.1\linewidth]{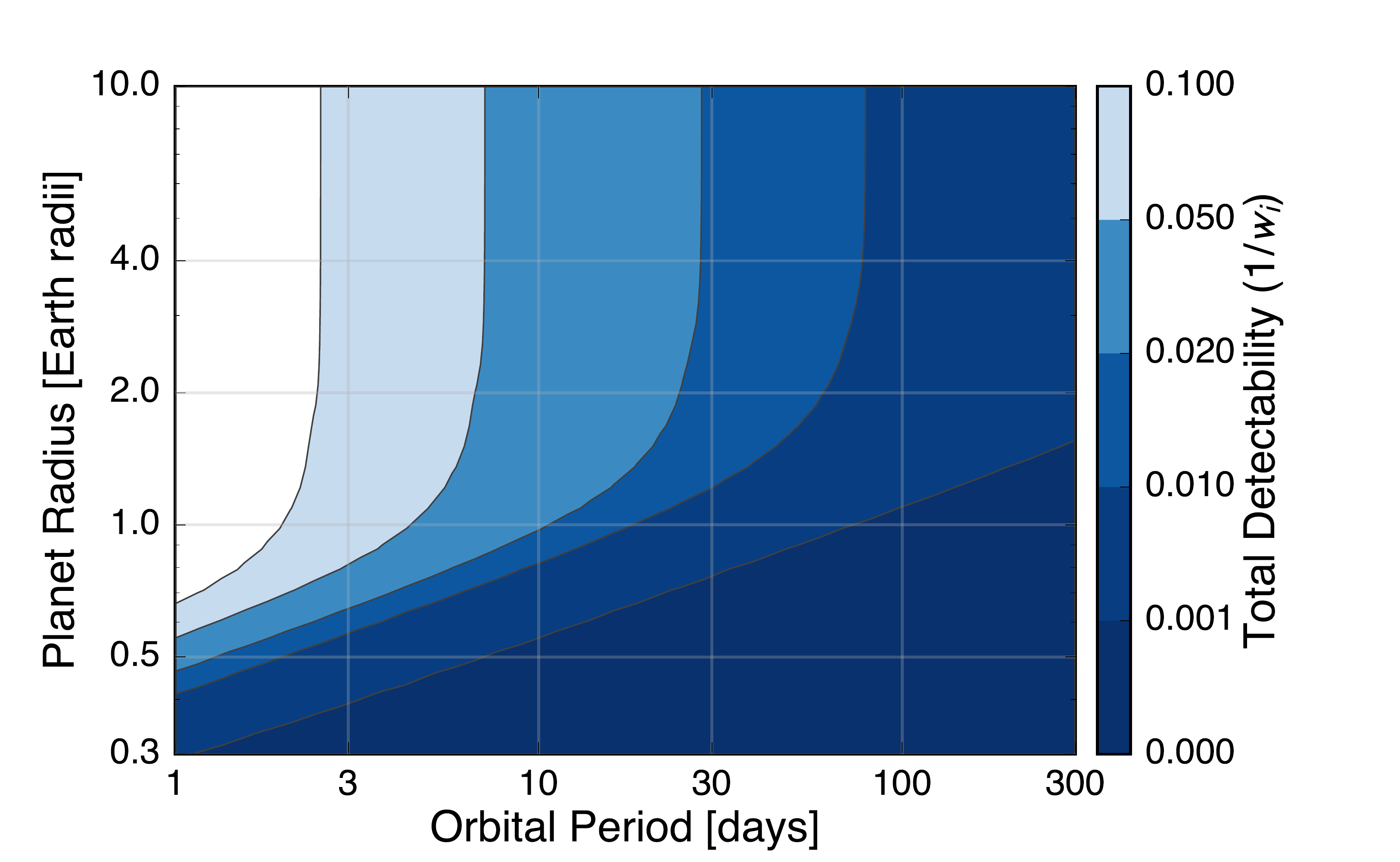}
\caption{\emph{Top:} Mean survey completeness for transiting planets orbiting the stars in our sample ($p_{\rm det}$).
\emph{Bottom:} Mean survey completeness for all planets orbiting stars in our sample ($p_{\rm det} \cdot p_{\rm tr}$).}
\label{fig:completeness}
\end{figure}

For each planet detection there are a number of similar planets that would not have been detected due to a lack of sensitivity or unfavorable geometric transit probability. To compensate, we weighted each planet detection by the inverse of these probabilities, 
\begin{equation}
w_i = \frac{1}{(p_{\rm det} \cdot p_{\rm tr})},
\label{eqn:weight}
\end{equation}
where $p_{\rm det}$ is the fraction of stars in our sample where a transiting planet with a given signal to noise ratio given by Equation \ref{eqn:snr} could be detected:
\begin{equation}
p_{\rm det} = \frac{1}{N_{\star}} \sum_i^{N_{\star}} C(m_i).
\end{equation}

The geometric transit probability is $p_{\rm tr} = 0.7R_{\star}/a$. The factor of 0.7 compensates for our omission of planet detections with $b > 0.7$ from the planet catalog. Figure \ref{fig:completeness} shows the mean pipeline completeness ($p_{\rm{det}}$) and mean total search completeness ($1/w_i$) as a function of planet radius and orbital period for the filtered Stellar17 sample of \Kepler target stars. The detection probabilites, transit probabilities, and weights ($w_i$) for each planet in our final catalog are listed in Table \ref{tab:weights}.

\subsection{Occurrence Calculation}
\label{sec:occurrence}

Following the definitions in \citet{Petigura13b}, the average planet occurrence rate (number of planets per star) for any discrete bin in planet radius or orbital period is the sum of these weights divided by the total number of stars in the sample ($N_{\star}$):
\begin{equation}
f_{\rm bin} = \frac{1}{N_{\star}}\sum_{i=1}^{n_{\rm pl,bin}} w_i.
\end{equation}
Again, $N_{\star}=\val{nstars-occ}$ is the total number of dwarf stars in the Stellar17 catalog that pass the same filters on stellar parameters that were applied to the planet catalog: no giant stars (selected using Equation \ref{eqn:giants}), 4700 K $<$ \teff $<$ 6500 K, and \Kp$ \leq 14.2$.

\section{The Planet Radius Gap}
\label{sec:valley}

\begin{figure*}[t]
\centering
\includegraphics[scale=0.48]{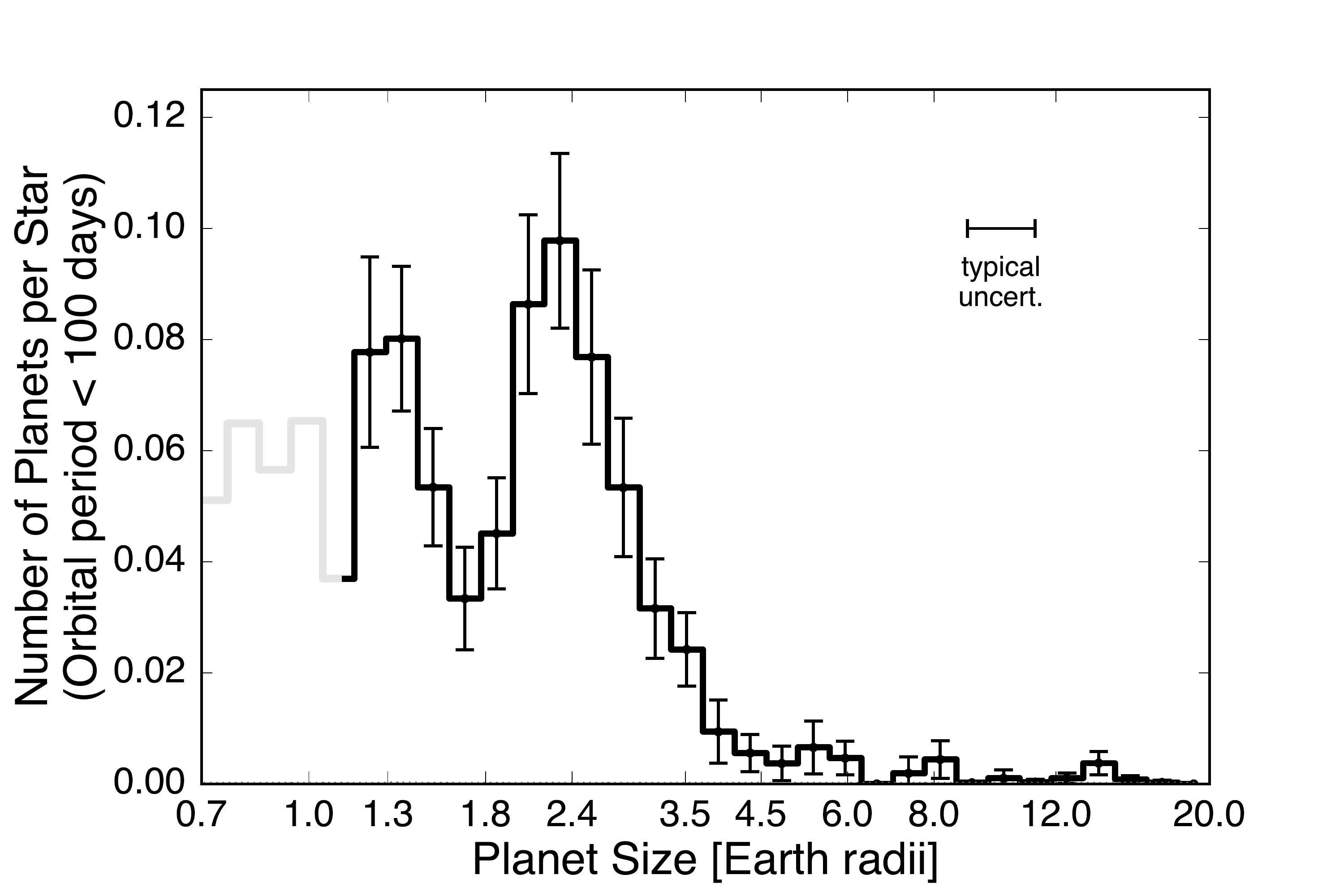}
\includegraphics[scale=0.48]{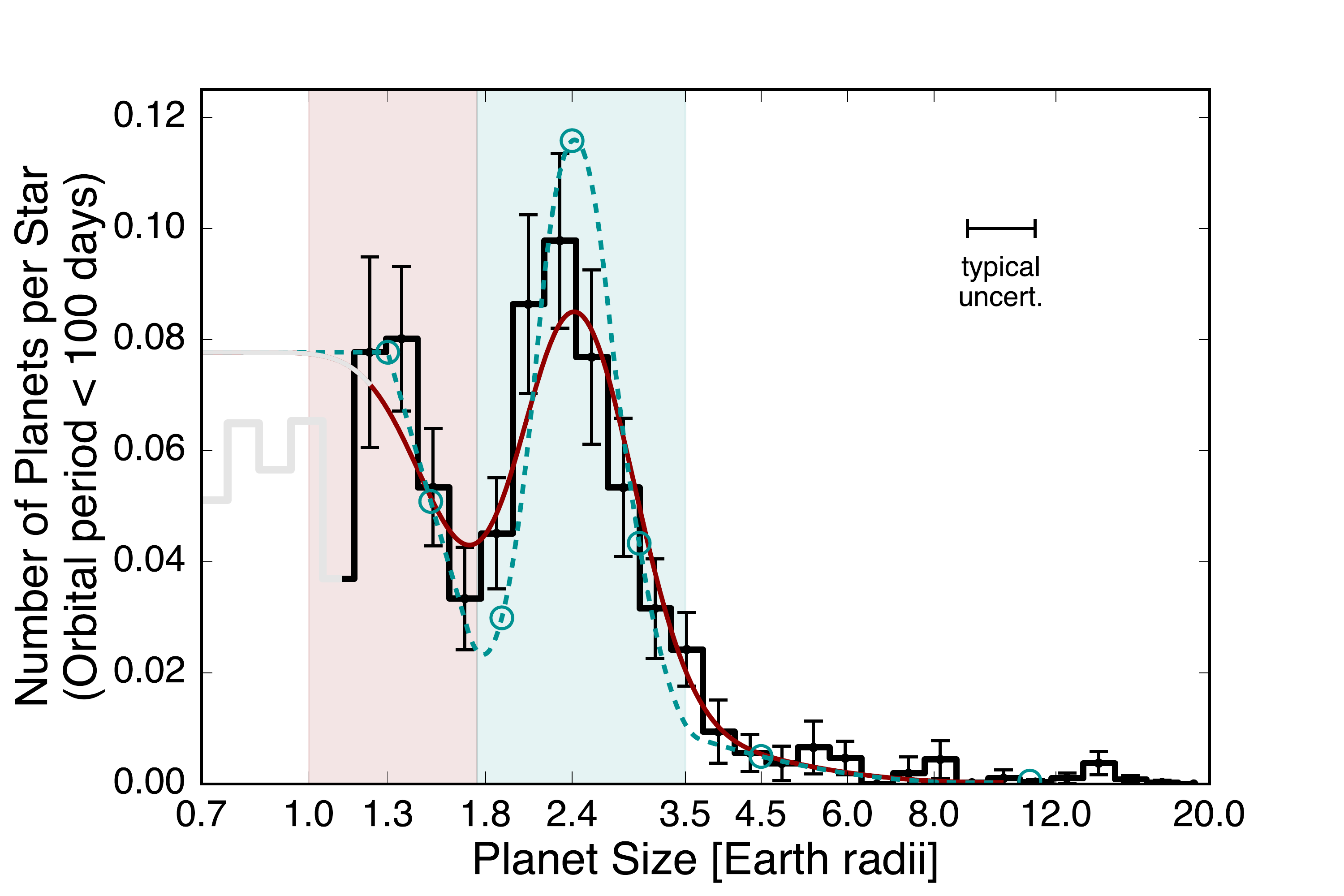}
\caption{
\emph{Top:} Completeness-corrected histogram of planet radii for planets with orbital periods shorter than 100 days. Uncertainties in the bin amplitudes are calculated using the suite of simulated surveys described in Section \ref{sec:validation}. The light gray region of the histogram for radii smaller than 1.14 \rearth suffers from low completeness. The histogram plotted in the dotted grey line is the same distribution of planet radii uncorrected for completeness. The median radius uncertainty is plotted in the upper right portion of the plot.
\emph{Bottom:} Same as top panel with the best-fit spline model over-plotted in the solid dark red line. The region of the histogram plotted in light grey is not included in the fit due to low completeness. Lightly shaded regions encompass our definitions of ``super-Earths'' (light red) and ``sub-Neptunes'' (light cyan). The dashed cyan line is a plausible model for the underlying occurrence distribution after removing the smearing caused by uncertainties on the planet radii measurements. The cyan circles on the dashed cyan line mark the node positions and values from the spline fit described in \S \ref{sec:spline}.}
\label{fig:valley}
\end{figure*}

Figure \ref{fig:valley} shows the completeness-corrected distribution of planet radii for the filtered sample of \val{num-planets-filtered} planets and the corresponding occurrence values are tabulated in Table \ref{tab:occ}. Uncertainties on the bin heights are calculated using Poisson statistics on the number of detections within the bin, scaled by the size of the completeness correction in each bin, and scaled again by a correction factor determined from a collection of simulated transit surveys as described in Section \ref{sec:validation}. The completeness corrections are generally small. We are sensitive to $>80$\% of 2.0 \rearth planets out to orbital periods of 100 days, and $>50$\% of 1.0 \rearth planets out to 30 days (Figure \ref{fig:completeness}). The transit probability term in Equation \ref{eqn:weight} dominates the corrections in most of the parameter space explored. Somewhat surprisingly, the larger, sub-Neptunes receive a completeness boost that is larger than the boost received by the smaller, super-Earths (compare the dotted grey line in Figure \ref{fig:valley} to the solid black line) because the sub-Neptunes tend to orbit at larger orbital distances where transit probabilities are smaller. The mean transit probability ($p_{\rm tr}$) for planets with radii of 1.0--1.75 \rearth in our sample is \val{tr-prob-sm} while the transit probability for planets with radii of 1.75--3.5 \rearth is a factor of two lower (\val{tr-prob-lg}). However, the mean detectability ($p_{\rm det}$) for those same two classes of planets are both very high at \val{det-prob-sm} and \val{det-prob-lg} respectively.

\begin{deluxetable}{lr}
\tabletypesize{\footnotesize}
\tablecaption{Planet Occurrence}
\tablehead{ 
    \colhead{Radius bin}   &   \colhead{Number of planets per star}  \\
    \colhead{\rearth}        &    \colhead{$f_{\rm bin}$ for $P < 100$ d} } 
\startdata
1.16--1.29  &  $0.078 \pm 0.017$  \\
1.29--1.43  &  $0.08 \pm 0.013$  \\
1.43--1.59  &  $0.053 \pm 0.011$  \\
1.59--1.77  &  $0.0334 \pm 0.0092$  \\
1.77--1.97  &  $0.05 \pm 0.01$  \\
1.97--2.19  &  $0.086 \pm 0.016$  \\
2.19--2.43  &  $0.098 \pm 0.016$  \\
2.43--2.70  &  $0.077 \pm 0.016$  \\
2.70--3.00  &  $0.053 \pm 0.012$  \\
3.00--3.33  &  $0.0316 \pm 0.0089$  \\
3.33--3.70  &  $0.0242 \pm 0.0066$  \\
3.70--4.12  &  $0.0094 \pm 0.0057$  \\
4.12--4.57  &  $0.0056 \pm 0.0034$  \\
4.57--5.08  &  $0.0037 \pm 0.0031$  \\
5.08--5.65  &  $0.0066 \pm 0.0048$  \\
5.65--6.27  &  $0.005 \pm 0.003$  \\
6.27--6.97  &  $0.0 \pm inf$  \\
6.97--7.75  &  $0.0019 \pm 0.0029$  \\
7.75--8.61  &  $0.0044 \pm 0.0034$  \\
8.61--9.56  &  $0.00022 \pm 0.00032$  \\
9.56--10.63  &  $0.001 \pm 0.0015$  \\
10.63--11.81  &  $0.00035 \pm 0.00053$  \\
11.81--13.12  &  $0.00104 \pm 0.00094$  \\
13.12--14.58  &  $0.0038 \pm 0.0021$  \\
14.58--16.20  &  $0.00084 \pm 0.00066$  \\
16.20--18.00  &  $0.0003 \pm 0.0004$  
\enddata
\vspace{10pt}

\label{tab:occ}
\end{deluxetable}

\subsection{Comparison with Log-Uniform Distribution}

We performed several tests to quantify the significance of the gap in the planet radius distribution. First, we performed a two-sided Kolmogorov-Smirnov \citep[K-S,][]{Kolmogorov33,Smirnov48} test to assess the probability that the planet radius number distribution for radii in the range 1--3 \rearth is drawn from a log-uniform distribution. This test returns a probability of \val{kstest} that the planet radii between 1--3 \rearth are drawn from a log-uniform distribution. However, we note that blind interpretation of p-values from K-S tests can often lead to overestimates of significance \citep{Babu06}. Similarly, an Anderson-Darling test also rejects the hypothesis that the planet radii between 1--3 \rearth were drawn from a log-uniform distribution with a p-value of \val{adtest}.

\subsection{Dip Test of Multimodality}

Hartigan's dip test is a statistical tool used to estimate the probability that a sample was drawn from a unimodal distribution or a multi-modal distribution with $\geq$2 modes \citep{Hartigan85}. It is similar to the K-S statistic in that it measures the maximum distance between an empirical distribution and a unimodal distribution. Applying this test to the number distribution of $\log{\Rp}$ for planet radii in the range 1--3 \Re returns a p-value of 1.4$\times10^{-3}$ that the distribution was drawn from a unimodal distribution. This strongly suggests that the planet radius distribution is multi-modal.

\subsection{Spline Model}
\label{sec:spline}

\begin{deluxetable}{lrr}
\tabletypesize{\small}
\tablecaption{Spline Fit}
\tablehead{ 
    \colhead{Node Location}   &   \colhead{Best-fit Value}               & \colhead{1 $\sigma$ Credible Interval} \\
    \colhead{\rearth}        &    \colhead{($f_{\rm bin}$)}  & \colhead{($f_{\rm bin}$)} 
}
\startdata
1.3  & 	0.078	&	fixed \\
1.5  &  	0.051		&	$0.05 \pm 0.02$ \\
1.9 &  	0.030		&	$0.03 \pm 0.02$ \\
2.4 &  	0.116		&	$0.11 \pm 0.01$ \\
3.0 &  	0.043	&	$0.044 \pm 0.005$ \\
4.5 &  	0.0050	&	$0.005 \pm 0.002$ \\
11.0 &	0.00050	&	$0.0005 \pm 0.0003$
\enddata
\vspace{10pt}

\label{tab:splines}
\end{deluxetable}

Modeling the planet radius distribution with splines having nodes at fixed values gives a good fit for a range of planet sizes.  Virtues of this model are the small number of free parameters and model flexibility, particularly in asymptotic regions where others models (e.g.\ Gaussians) force the distribution to zero. We fit a second-order spline with seven node points fixed at specific radii to the weighted histogram of planet occurrence. We excluded from the fit bins for radii smaller than 1.14 \rearth where the pipeline completeness at $P=100$ days is less than 25\%. The model was adjusted by varying the amplitudes of the spline nodes, then convolving with a Gaussian kernel whose width is the median fractional planet radius uncertainty (\val{cks-rp-frac-err-median}). The convolved model is averaged over each of the histogram bins before performing the $\chi^2$ comparison. This allows us to separate the smearing of the observed distribution due to measurement uncertainties from a ``deconvolved'' view of the underlying distribution. Again we found the best-fit solution using the \texttt{lmfit} package to minimize the normalized residuals of the histogram bins relative to the convolved model. We used the \texttt{emcee} \citep{DFM13} interface built into \texttt{lmfit} to estimate the uncertainties on the node values. We performed the fits working in $\log(R_{P})$ and required positive occurrence values for the deconvolved model. For radii outside of the range spanned by our node locations, we extrapolated assuming constant (log-uniform) occurrence.

Deconvolution is an inherently unstable process and we caution against over-interpretation of the deconvolved model. Our best-fit deconvolved model is not the only solution that could produce an equivalent convolved model. The deconvolved model is also somewhat sensitive to the choice of the node locations, while the convolved model is insensitive to those choices. However, the deconvolved model suggests that the gap is likely deeper than observed. This motivates detailed follow-up and characterization of the planets that fall within the gap. The best-fit model (red line) and deconvolved model (dashed cyan line) are both over-plotted on the completeness-corrected planet radius distribution in Figure \ref{fig:valley}. Table \ref{tab:splines} lists the locations, best-fit values, and 1$\sigma$ credible intervals for the spline nodes.

\subsection{Relative Frequency of Super-Earths and Sub-Neptunes}
\label{sec:rel_occ}

Many authors use the terms ``super-Earth'' and ``sub-Neptune'' interchangeably, or draw arbitrary distinctions in mass or radius between these two classes. The observed gap in the radius distribution of small planets suggests a less arbitrary division.  In the text below we define a ``super-Earth'' as a planet with a radius of 1-1.75 \rearth, and a ``sub-Neptune'' as having a radius of 1.75--3.5 \rearthe. 

We calculated the occurrence ratio of super-Earths to sub-Neptunes to be \val{occ-ratio}. The uncertainty is determined using a suite of simulated surveys described in Appendix \ref{sec:validation}. The nearly equal occurrence of super-Earths and sub-Neptunes with $P < 100$ days provides an important constraint for planet formation models. This is likely a lower limit on this ratio since the super-Earth domain likely extends to sizes smaller than 1.1 \rearthe. 

\subsection{Two-Dimensional Weighted Kernel Density Estimation}

\begin{figure*}[ht]
\centering
\includegraphics[width=0.95\linewidth]{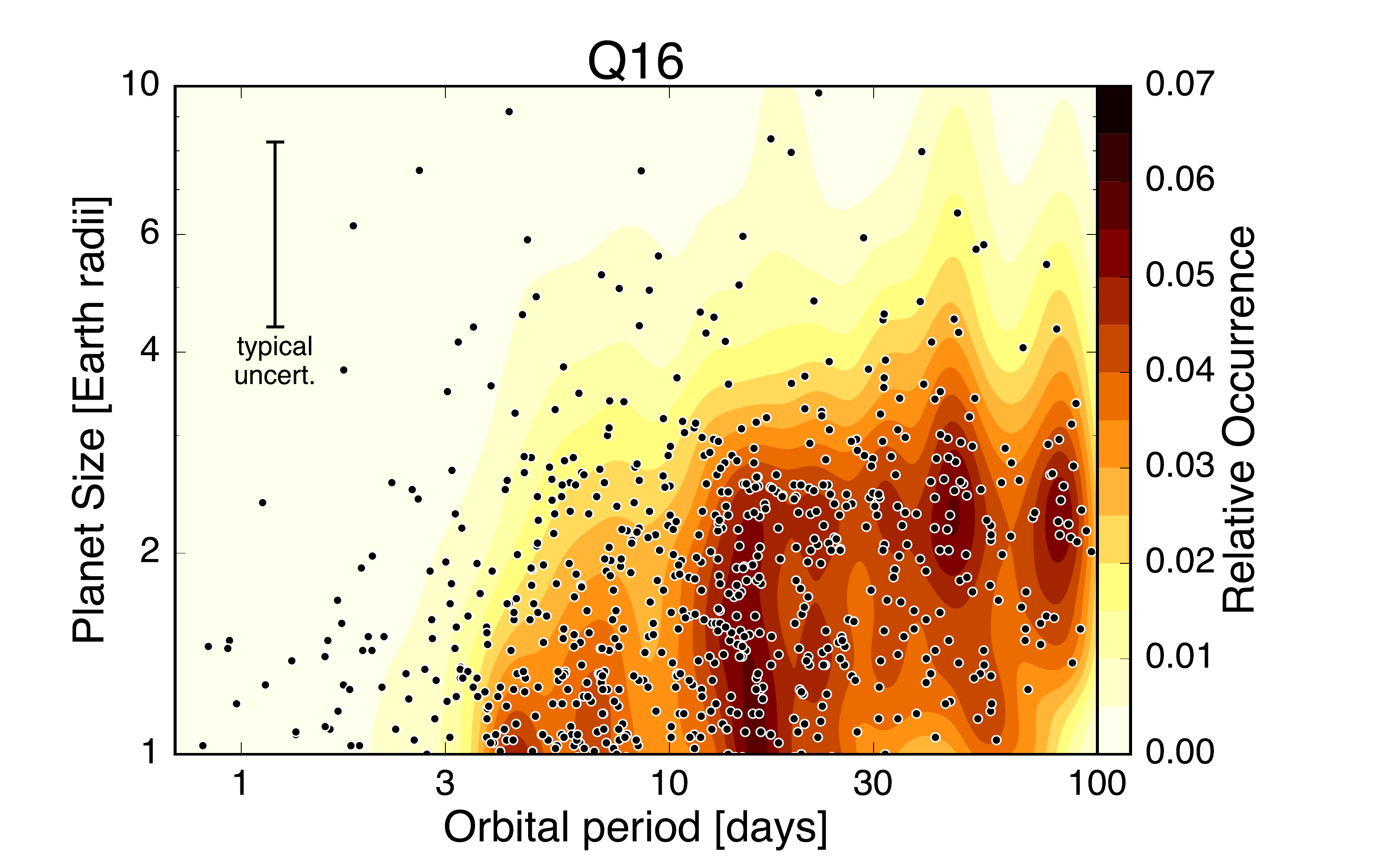}
\includegraphics[width=0.95\linewidth]{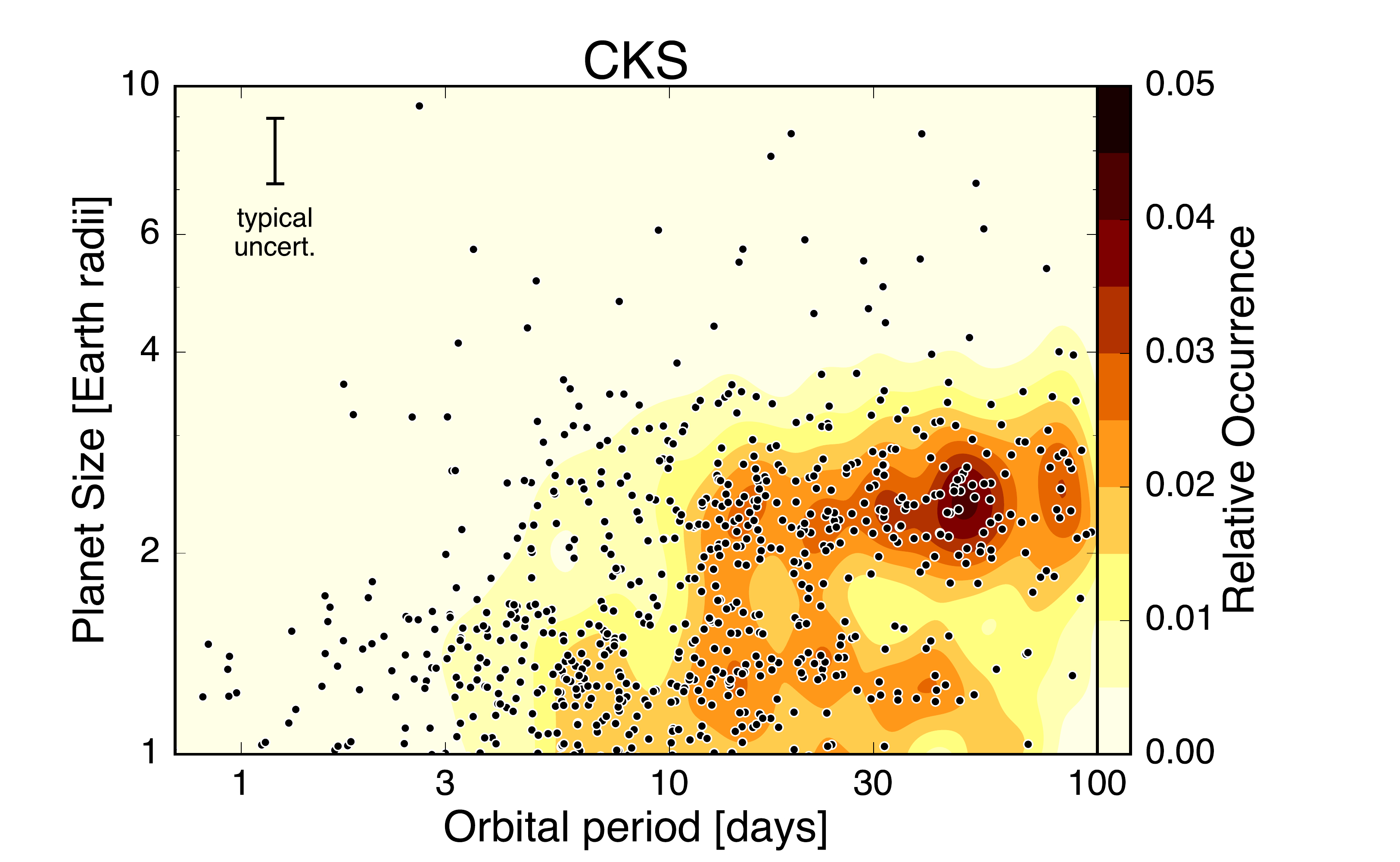}
\caption{\emph{Top:} Two-dimensional planet radius distribution as a function of orbital period using stellar parameters from the Q16 catalog. \emph{Bottom:} Two-dimensional planet radius distribution as a function of orbital period using updated planet parameters from Paper II. In both cases the median uncertainty is plotted in the upper left. Individual planet detections are plotted as black points. The contours are corrected for completeness using the wKDE technique. }
\label{fig:per}
\end{figure*}

\begin{figure*}[ht]
\centering
\includegraphics[width=0.95\linewidth]{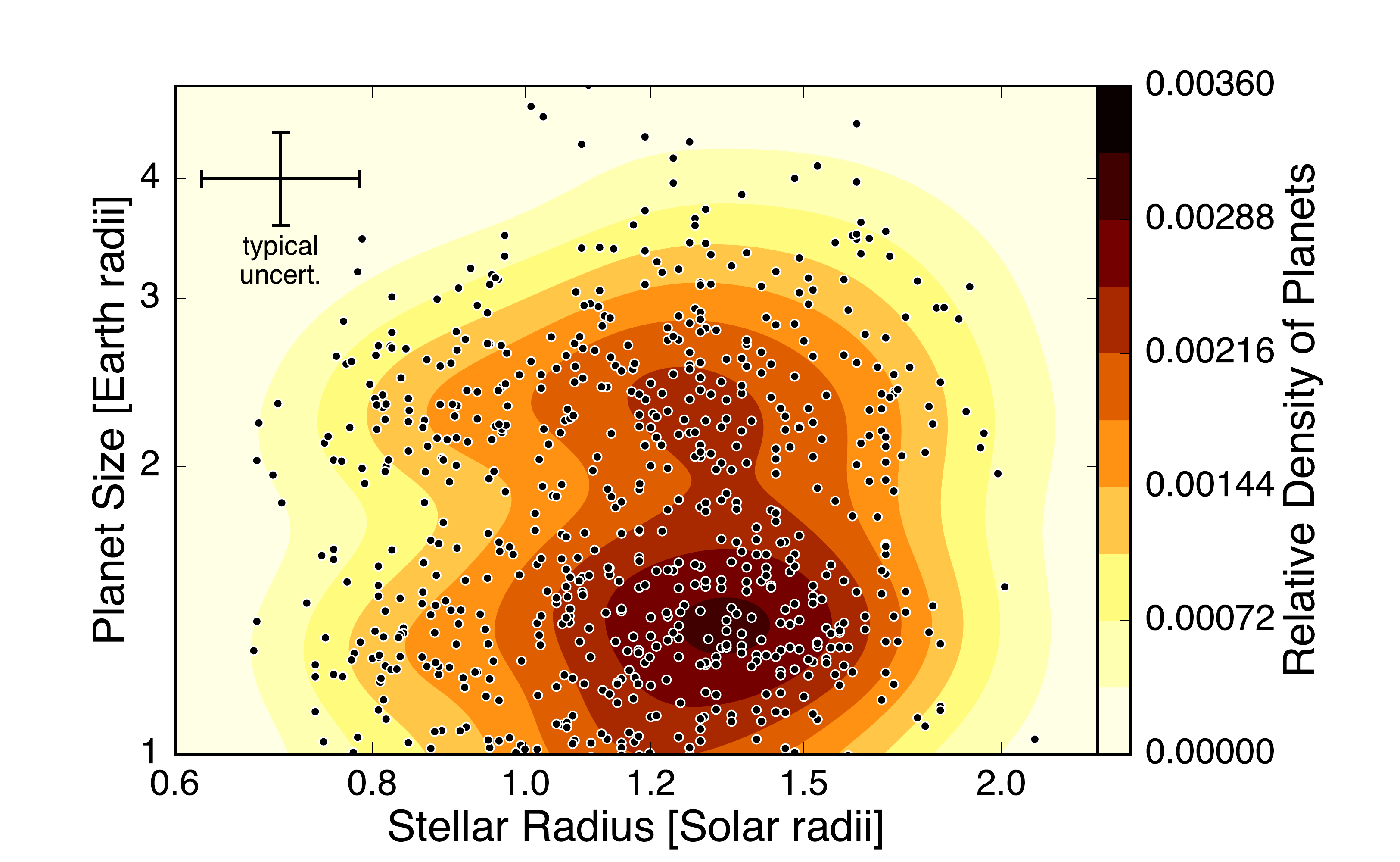}
\caption{Two-dimensional planet radius distribution as a function of stellar radius using updated planet parameters from Paper II. The median uncertainty is plotted in the upper left. Individual planet detections are plotted as black points. The underlying contours are not corrected for completeness. The bifurcation of planet radii is independent of the size of the host star.}
\label{fig:rstar}
\end{figure*}

\begin{figure*}[ht]
\centering
\includegraphics[width=0.95\linewidth]{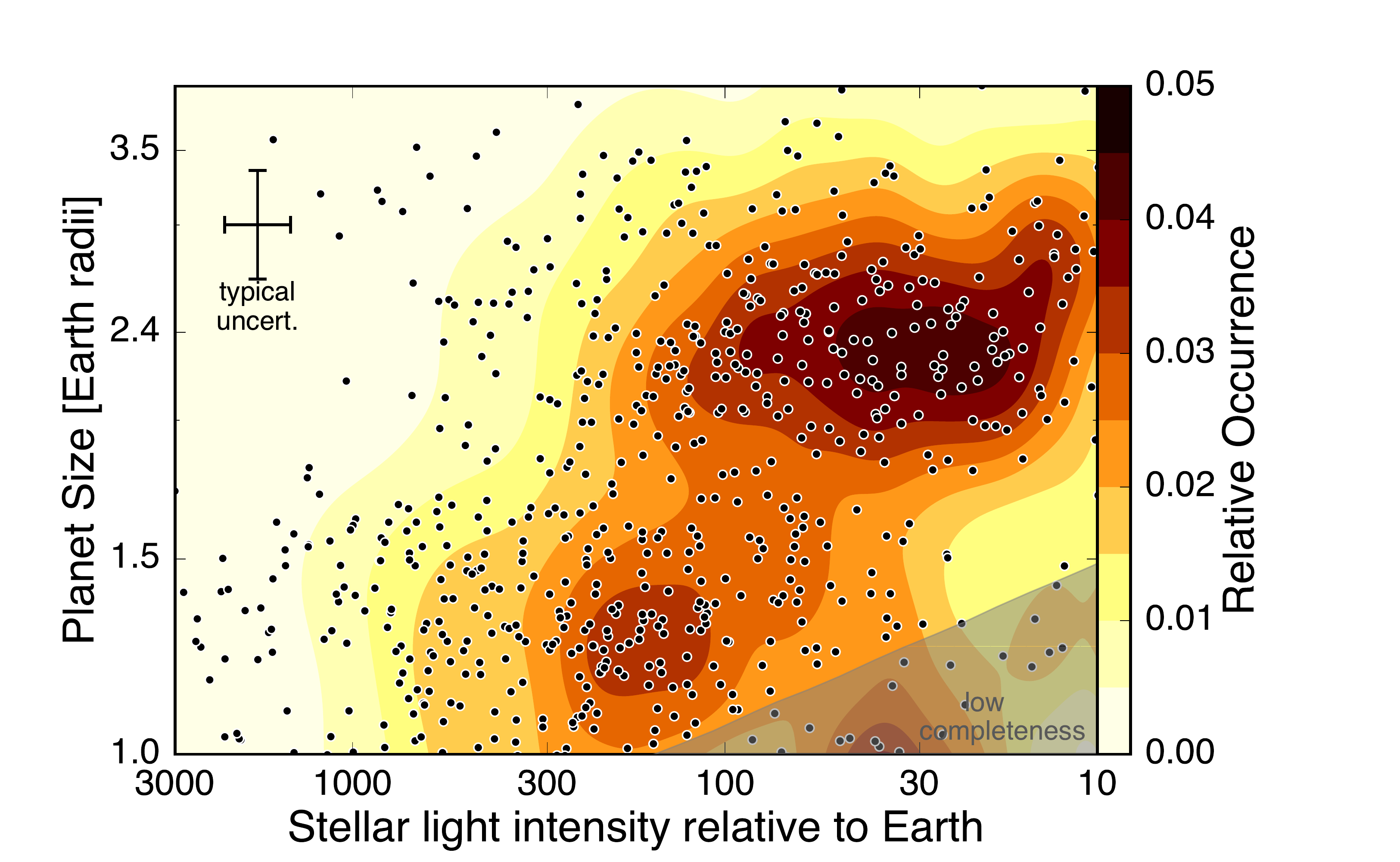}
\includegraphics[width=0.95\linewidth]{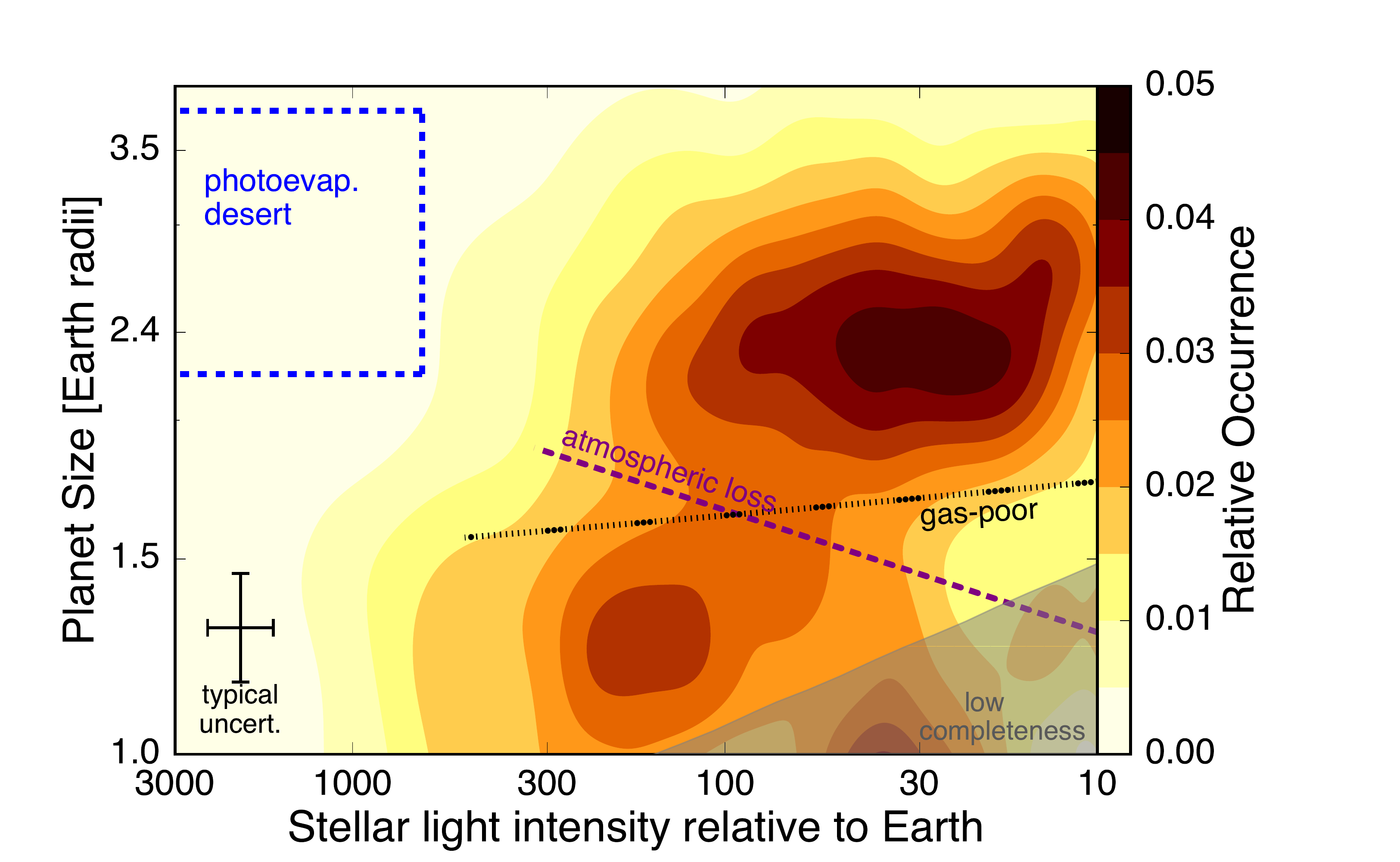}

\caption{
\emph{Top:} Two-dimensional distribution of planet size and incident stellar flux. The median uncertainty is plotted in the upper left. There are at least two peaks in the distribution. One class of planets has typical radii of $\sim$1.3 \rearth and generally orbit in environments with \Sinc $>$100 \Searthe, while another class of slightly larger planets with typical radii of $\sim$2.4 \rearth orbit in less irradiated environments with \Sinc $< 200$ \Searthe. 
\emph{Bottom:} Same as top panel with individual planet detection points removed, annotations added, and vertical axis scaling changed. The region enclosed by the dashed blue lines marks the photoevaporation desert, or hot-Super Earth desert as defined by \citet{Lundkvist16}. The shaded region in the lower right indicates low completeness. Pipeline completeness in this region is less than 25\%. The purple and black lines show the scaling relations for the photoevaproation valley predicted by \citet{Lopez16} for scenarios where these planets are the remnant cores of photoevaporated Neptune size planets (dashed purple line) or that these planets are formed at late times in a gas-poor disk (dotted black line).}
\label{fig:sinc}
\end{figure*}

In the following subsections we present and discuss several contour plots. The contours were derived using the Weighted Kernel Density Estimation (wKDE) technique described in Appendix \ref{sec:wKDE} and have all been corrected for completeness (with the exception of Figure \ref{fig:rstar}). We calculated bi-variate Gaussians for each pair of planet parameters over a fixed high-resolution grid in the two parameters, sum these Gaussians over all planets, and divide by the total number of stars in the sample ($N_{\star}$=\val{nstars-occ}). Each bi-variate Gaussian is normalized to have a maximum value of 1.0, then multiplied by the weight associated with the given planet detection ($w_i$, Equation \ref{eqn:weight}). The points plotted are the CKS parameters.

\subsubsection{Planet Radius vs.\ Orbital Period}

We first look at the distribution of planet radii as a function of orbital period ($P$). Figure \ref{fig:per} shows the distribution of planet radii as a function of orbital period for planet and stellar parameters from the Q16 catalog (\emph{top panel}).  It also offers a comparison with the same distribution derived from the CKS parameters (\emph{bottom panel}).

There is a declining number of small planet detections going toward long orbital periods. However the underlying completeness-corrected contours suggest that the occurrence rate of these planets does not fall off with the number of detections. Instead, the lack of detections is likely an artifact of decreasing transit detectability and probability.

Figure \ref{fig:per} shows that small planets are significantly more common than large planets. The fact that planets smaller than Neptune (4 \rearthe) are much more common than Jovian-size planets has been well documented in the literature (e.g. \citet{Howard10, Mayor11, Howard12, Fressin13, Dong13, Petigura13b, Dressing15, Burke15}). However, the increase in  occurrence with decreasing planet size is evidently more rapid than was apparent in previous studies.

There is another feature in the \Rp vs. $P$ occurrence distribution that motivates a closer examination of the planet radius distribution along other axes. There are very few planets larger than 2 \rearth with orbital periods shorter than about 10 days while planets with radii smaller than 1.8 \rearth remain quite common down to orbital periods of about 3 days. A sharp decline in the occurrence rate of planets larger than approximately 1.6 \rearth with orbital periods shorter than 10 days has been previously observed \citep{Howard12, Dong13, Sanchis-Ojeda14}.

\subsubsection{Planet Radius vs.\ Stellar Radius}

Figure \ref{fig:rstar} shows the distribution of planet size as a function of host star size. This distribution shows two distinct populations of planets with a gap separating them. Planets appear to preferentially fall into two classes, one with radii of $\sim$2.4 \rearth and another with radii of $\sim$1.3 \rearthe. Planets with intermediate radii of 1.5-2.0 \rearth are comparatively rare. The gap occurs at the same planet radius for all stellar sizes in our sample. The bimodal planet size distribution holds true for planets orbiting stars with radii ranging from 0.7 \Rsun to 2.0 \Rsun.

\subsubsection{Planet Radius vs.\ Incident Flux}

\begin{figure}
\centering
\includegraphics[scale=0.37]{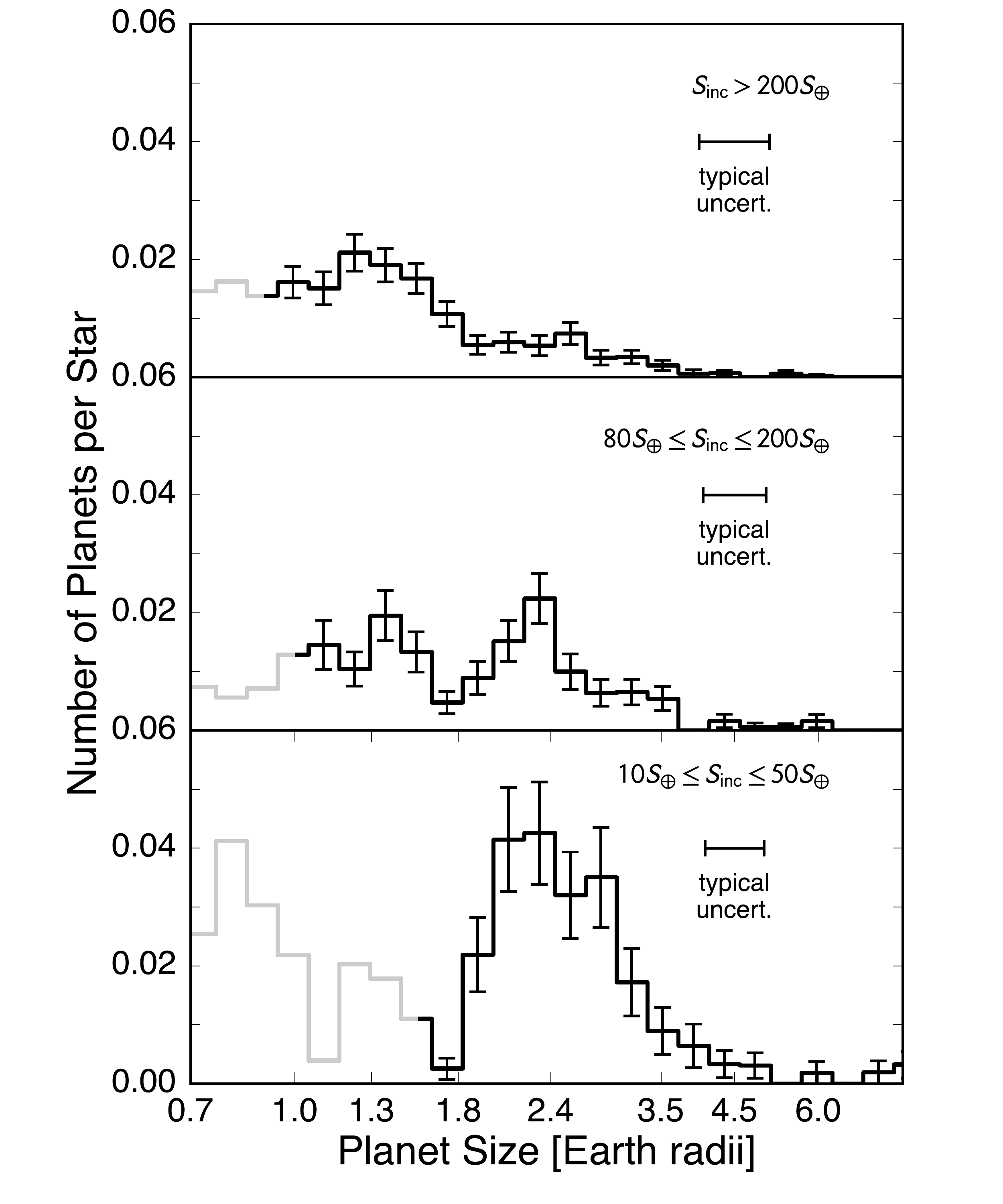}
\caption{Histograms of planet radii broken up into the ranges of incident flux (\Sinc) annotated in the upper right region of each panel. Planets orbiting in environments of higher \Sinc tend to be smaller than those in low \Sinc environments. Regions of the histograms plotted in light grey are highly uncertain due to pipeline completeness ($<$25\%).}
\label{fig:temphists}
\end{figure}

Figure \ref{fig:sinc} shows the planet radius distribution as a function of incident flux. The two planet populations shear apart in this domain. There is a dearth of sub-Neptunes orbiting in high incident flux environments. This trend is also visible in one-dimensional histograms of planet radii when broken up into groups based on \Sinc (Figure \ref{fig:temphists}). Most of the planets that contribute to the peak in the marginalized radius distribution at 1.3 \rearth are orbiting in environments with $\Sinc > 200$ \Searthe, while the planets that contribute to the peak at 2.4 \rearth experience $\Sinc < 80$ \Searthe. It is clear that the gap is present even at low incident fluxes and the two-dimensional \Sinc and period distributions show a potential deepening and/or widening of the gap toward lower incident fluxes. However, we can not determine if the gap radius is dependent on incident flux, or if the break radius is constant as a function of incident flux due to lack of completeness for small planets orbiting in cool environments.

There is also an upper envelope of planet size which decreases as a function of incident flux. Although there are a few exceptions, there is a clear dearth of planets in the upper left quadrant of Figure \ref{fig:sinc}. These should be some of the easiest planets to detect yet they do not appear in our sample of planets. This feature has been previously observed \citep[e.g.,][]{Howard12, Mazeh16, Lundkvist16} but our larger sample of planets with high-precision host star properties sharpens the boundary. The lack of planet detections in the lower right region of Figure \ref{fig:sinc} is the result of low survey completeness for small, long-period planets.

\section{Discussion}
\label{sec:discussion}

\begin{figure}
\centering
\includegraphics[scale=0.3]{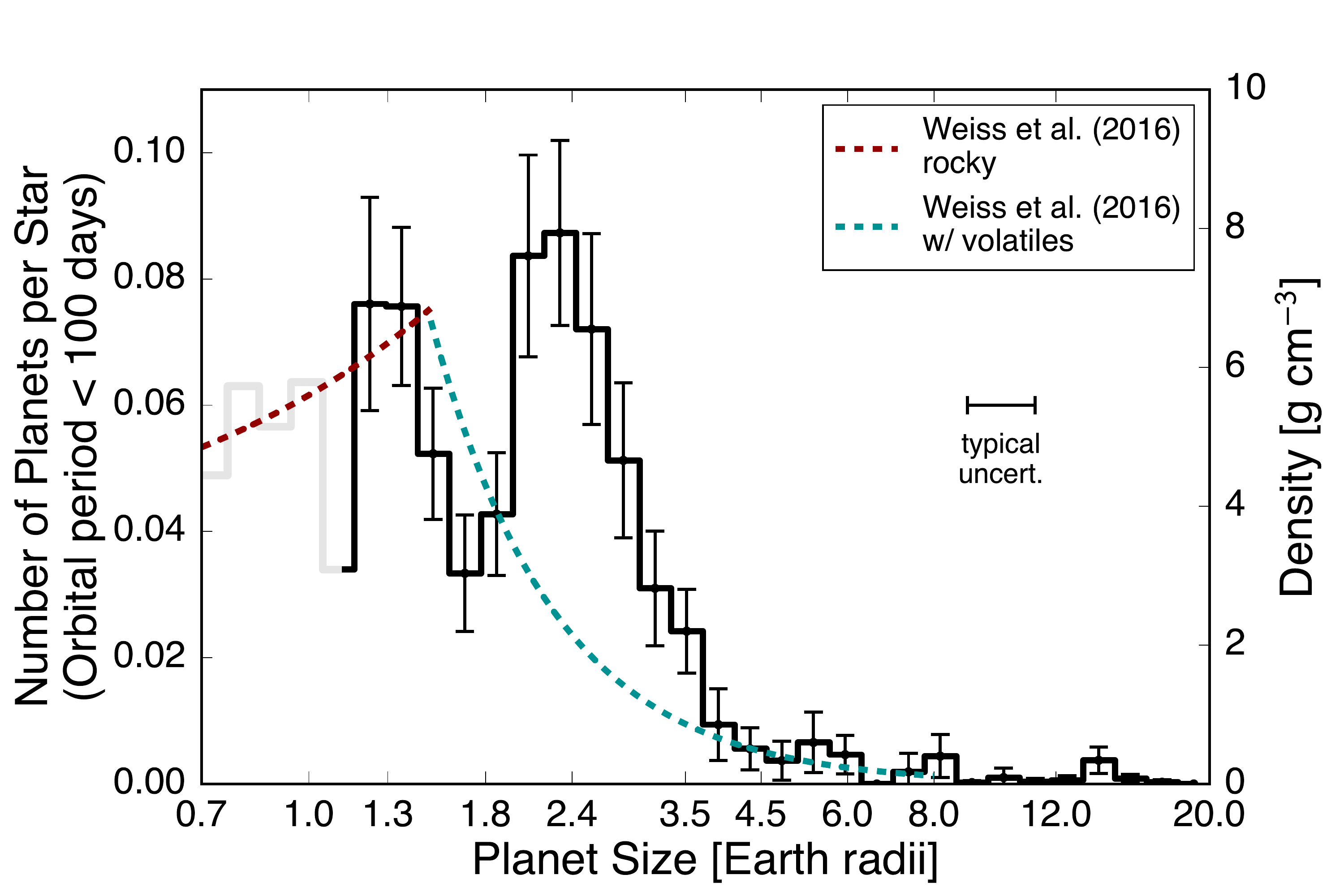}
\caption{An empirical fit to planet radius and mass measurements from \citep{Weiss17} over-plotted on the completeness-corrected planet radius distribution derived in this work. The maximum in the planet density fit peaks near the gap in the planet radius distribution.}
\label{fig:density}
\end{figure}

\begin{figure}
\centering
\includegraphics[scale=0.3]{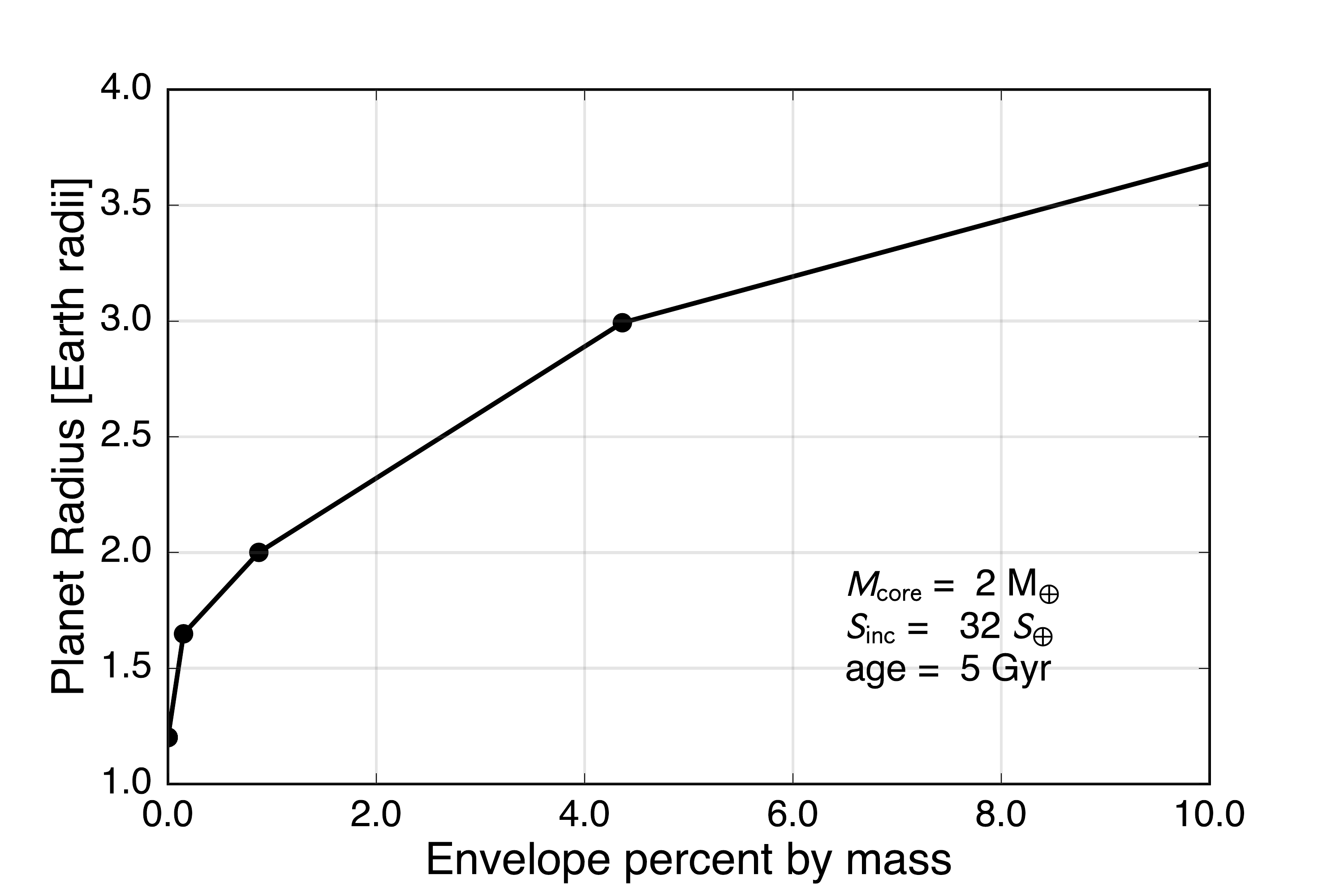}
\caption{Model for planet radius as a function of envelope size from \citet{Lopez13}. The final planet radius is plotted for a simulated planet with a 2 \mearth core mass that has been irradiated by 32 times the incident flux received by Earth for a period of 5 Gyr. A bare 2 \mearth core has a radius of 1.2 \rearthe. Adding an envelope of H/He which is less than 0.2\% of the planet's mass inflates the planet to over 1.6 \rearthe. An additional 0.7\% envelope by mass inflates the planet to 2 \rearthe.}
\label{fig:env}
\end{figure}

We have provided observational evidence that the distribution of planet sizes is not smooth (Figure \ref{fig:valley}).  Small planets have characteristic sizes of $\sim$1.3 \rearth (super-Earths) and $\sim$2.4 \rearth (sub-Neptunes).  These two planet populations each have intrinsic widths in their size distributions, but there is a gap that separates them.  Intermediate-size planets with radii of $\sim$1.5--2.0 \rearth are comparatively rare.

\subsection{Previous Studies of the Radius Distribution}

Many studies have examined the planet radius distribution using the \Kepler sample.  To date, none have shown statistically significant evidence for a gap in the distribution at 1.5--2.0 \rearthe.

The pioneering study of \citet{Owen13} pointed out a marginally-significant gap at $\sim$1.5--2 \rearth in the observed radius distribution and interpreted it as connected to the high-energy irradiation history of the planets.
They did not have a large set of accurate planet radii and they did not perform the completeness corrections necessary to confirm the feature. 
Here, we firmly detect a gap in the planet radius distribution between two peaks at 2.4 \rearth and $\leq1.3$ \rearth.

Based on the initial \Kepler planet catalog, \cite{Howard12} investigated
the domain of planets with \Rp $>$ 2\rearth and $P < 50$ days. They demonstrated that small planets are common.  However, they did not examine the detailed shape of the small planet occurrence function, due to the severe lack of completeness to small planets with the early \Kepler data releases, and large uncertainties in the planetary radii. At that time, the planetary radii were based on the relatively coarse estimates of the stellar radii from the KIC.

Follow-on studies \citep{Youdin11,Catanzarite11,Traub12} were similarly limited.
\cite{Dong13} benefited from a larger dataset. They focused on the orbital period distribution, with large (factor of two) bins in planet radius.
\citet{Petigura13b} utilized a much longer photometric time series (lasting 15 of 17 \Kepler quarters), and a custom planet detection pipeline enabling completeness corrections, but the sample was only large enough to allow for three bins in the radius range 1.0--2.8 \rearthe. 
\citet{Silburt15} measured occurrence for planets with radii between 1.0 and 4.0 \rearth and orbital periods between 20 and 200 days. They found a peak in the distribution near 2.4 \rearth and a slight decline in the frequency of smaller planets.
More recently, \citet{Burke15} studied the occurrence of small, long-period planets.  With 1$\sigma$ significance, they observed a diminution in planet occurrence in the 1.5--2.0 \rearth interval for planets having $P = 300$--700 days.

\subsubsection{Occurrence Rate Comparisons}
\label{sec:occ_comp}
Table 5 compares the occurrence rates measured in this work to those of several touchstone studies from the literature: \citet[][H12]{Howard12}, \citet[][P13]{Petigura13a}, \citet[][F13]{Fressin13}, and \citet[][M15]{Mulders15}. These works all analyzed Kepler planets, but used catalogs constructed from different amounts of Kepler photometry. In addition, these studies applied different treatments of pipeline completeness, adopted different false positive rates, analyzed different sub-samples of Kepler stars, and accounted for multi-planet systems in different ways. All of these differences can significantly affect the derived occurrence \citep{Burke15}. However, the relative occurrence rates between bins are insensitive to most of these issues and potential discrepancies in the absolute occurrence rates do not affect the presence or shape of the gap in the radius distribution. 

We choose to closely compare our occurrence values in this work to those of P13, because they used a nearly complete photometric dataset (43/48 months)
\footnote{
H12, F13, P13, and M15 used 4, 16, 43, and 22 months of photometry, respectively.
}
and corrected for pipeline completeness through direct injection and recovery. Our occurrence rates are typically 50\% higher than those of P13. However, P13 (and H12) measured the fraction of stars with planets as opposed to the number of stars per planet measured in this work (and in F13 and M15). The number of planets per star will always be larger than the fraction of stars with planets due to multi-planet systems. P13 estimated that their occurrence rates would have been 25--45\% higher if they had included multi-planet systems (depending on period and radius limits), which can reconcile much of differences between the two studies.

In comparing to previous results, we find that 2--2.8 \rearth are more common than 1.4--2 \rearth planets, in a relative sense. For example, we find that P13 found that 18.6\% of stars had a 2--2.8 \rearth planet with P $<$ 100 d vs. 14.2\% of stars with a 1.4--2 \rearth planet in the same period range. This corresponds to a ratio of 18.6/14.2 = 1.3. In this work, that ratio is 16.1\% / 27.0\% = 0.6. We can understand this difference in terms of the gap between 1.5 and 2.0 \rearth and the peak between 2.0 and 2.4 \rearth that emerged after we refined the host star radii through spectroscopy. Planets with true sizes between 2.0 and 2.8 \rearth were often scattered to the 1.4--2.0 \rearth bin due to the 40\% radius uncertainties from photometry. Thus the peak from 2.0--2.8 \rearth was diminished, while the gap from 1.4--2.0 \rearth was filled in. In summary, the integrated occurrence rates presented are largely consistent with previous works, with differences in the detailed radius distribution, owing to improved stellar radii.

\begin{deluxetable*}{llccccc}
\tabletypesize{\footnotesize}
\tablecaption{Occurrence Rate Comparison}
\tablehead{
\colhead{Radius Interval}   &   \colhead{Period Interval}   & \colhead{{\bf This Work}\tablenotemark{1}}  & \colhead{H12\tablenotemark{2,6,7}} & \colhead{P13\tablenotemark{3,6}} & \colhead{F13\tablenotemark{4}} & \colhead{M15\tablenotemark{5}} \\
\colhead{\rearth}        &    \colhead{(days)}  &  \colhead{($f_{\rm bin}$ \%)}   & \colhead{($f_{\rm bin}$ \%)}   & \colhead{($f_{\rm bin}$ \%)}   &  \colhead{($f_{\rm bin}$ \%)}  &  \colhead{($f_{\rm bin}$ \%)}
}
\startdata
1.4--2.8  & $<100$  	&	$43.1\pm2.2$   &  \nodata             &  $32.8\pm1.4$   &   $35.0\pm2.8$\tablenotemark{8} & $26.7\pm1.7$\tablenotemark{8}  \\
2--2.8  & 	$<50$  	&	$19.4\pm1.4$   &  $9.0\pm1.5$     &  $18.6\pm1.6$   &   $17.5\pm1.6$                             & $12.8\pm0.5$  \\
2--4  & 	$<50$  	&	$25.4\pm1.6$   &  $13.0\pm0.8$   &  $16.6\pm1.8$   &   $18.3\pm1.3$                             & $18.6\pm0.6$  \\
2--4  & 	$<100$  	&	$36.6\pm2.2$   &  \nodata             &  $24.1\pm2.3$   &   $24.0\pm2.2$\tablenotemark{8} & $22.9\pm0.8$\tablenotemark{8} 
\enddata
\vspace{10pt}

\tablenotetext{1}{Uncertainties do not include the scaling factors derived in Appendix \ref{sec:validation}}
\tablenotetext{2}{\citet{Howard12}}
\tablenotetext{3}{\citet{Petigura13a}}
\tablenotetext{4}{\citet{Fressin13}}
\tablenotetext{5}{\citet{Mulders15}}
\tablenotetext{6}{Measured fraction of stars with planets instead of number of planets per star}
\tablenotetext{7}{Only studied planets with periods shorter than 50 days and larger than 2 \rearth}
\tablenotetext{8}{Periods shorter than 85 days}
\tablecomments{Each occurrence rate study focused on different stellar samples, planet detection pipelines, period limits, etc.
This table is not meant to be an exact comparison of the results from each study, but instead a rough comparison to show
 general agreement or highlight large disagreements.}

\label{tab:occ_comp}
\end{deluxetable*}

\subsection{Rocky to Gaseous Transition}
Studies of the relationship between planet density and radius suggest that planet core sizes reach a maximum of about 1.6 \rearthe. Planets with larger radii and measured masses are mostly low-density and require an extended atmosphere to simultaneously explain their masses and radii \citep{Marcy14, Weiss14, Rogers15, Wolfgang15}. Figure \ref{fig:density} shows the radius distribution derived in this work and an empirical fit to the densities and radii of small planets \citep{Weiss17}. This fit to a sample of planets with measured densities peaks near our observed gap in the planet radius distribution. This suggests that the majority of planets smaller than the minimum in the occurrence distribution are rocky while larger planets likely contain enough volatiles to contribute significantly to the planets' radii.

Additionally, ultra-short-period planets (USPs, having $P < 1$ day) present a clean sample of stripped, rocky planet cores.  It is unlikely that H/He atmospheres could survive on small planets bathed in the intense irradiation experienced by USPs. These planets must be bare, rocky cores, stripped of any significant atmosphere. \citet{Sanchis-Ojeda14} found that the occurrence of ultra-short period planets falls off sharply for $R_P>1.6$ \rearthe.
The apparent lack of rocky cores larger than 1.6 \rearth also suggests that planets larger than that must have non-negligible volatile envelopes.

\subsection{Potential Explanations for the Gap}
\subsubsection{Photoevaporation}

Photoevaporation provides a possible mechanism to produce a gap in the radius distribution, even if the initial radius distribution was continuous \citep{Owen13}. 
\citet{Lopez16} modeled the masses and radii of planets with various gas envelope fractions.  A bare, rocky planet (no envelope) with a mass of 2 \mearth has a radius of 1.2 \rearth in their models.  Adding an H/He envelope with a mass of 0.002 \mearth (0.1\% mass fraction) increases the planet size to 1.5 \rearthe, a large change in size for a small change in mass.  Adding an additional 0.7\% by mass of H/He swells the planet to 2.0 \rearthe (see Figure \ref{fig:env}).  This non-linear mass-radius dependence on volatile fraction has two effects.  First, making a planet with a thin atmosphere requires a finely tuned amount of H/He.  Second, photoevaporating a planet's envelope significantly changes its size.  Our observation of two peaks in the planet size distribution is consistent with super-Earths being rocky planets with atmospheres that contribute negligibly to their size, while sub-Neptunes are planets that retain envelopes with mass fractions of a few percent.

\subsubsection{Gas-poor formation}
Accretion of a modest gas envelope poses a theoretical challenge because fine-tuning is required to end up with an appreciable atmosphere that does not trigger runaway gas accretion and giant planet formation.
\citet{Lee14} proposed a mechanism that produces small planets with low envelope fractions by delaying gas accretion until the gas in the protoplanetary disk is nearly dissipated. They also proposed that small planets could form in very metal-rich disks where high opacity slows cooling and accretion. 

In addition, a few-percent-by-mass secondary atmosphere can be outgassed during planet formation and evolution \citep{Adams08}. Our observed gap in the planet radius distribution could be explained by a mechanism that causes the creation of a secondary atmosphere during the formation of only $\sim$50\% of terrestrial planets.

\subsubsection{Impact Erosion}
Impacts can also provide a way to sculpt the atmospheric properties of small planets and strip large primordial envelopes down to a few percent by mass \citep[e.g.,][]{Schlichting15, Liu15, Inamdar16}. It is unclear whether a gap in the radius distribution could arise from impacts alone since impact erosion is a highly stochastic process. However, the atmospheric heating initiated by an impact can cause the envelope to expand, making it more susceptible to photoevaporation.

\subsubsection{Signatures of Atmospheric Sculpting}
\citet{Lopez16} considered two scenarios for the formation of sub-Neptunes/super-Earths. In one scenario, super-Earths are the remnant cores of photoevaporated, Neptune-size planets. In the other scenario super-Earths form late in the evolution of the protoplanetary disk, just as the gas dissipates \citep{Lee14}. They predict that the transition radius between these two populations (the gap that we observed) should be a function of semi-major axis. If super-Earths are evaporated cores then the transition radius should be larger at lower incident flux. However, if super-Earths form in a gas-poor disk, or lose gas during the late stages of formation due to giant impacts, then the transition radius should decrease with increasing orbital distance. The distribution of planet radii as a function of insolation flux (Figure \ref{fig:sinc}) does not show a clearly increasing or decreasing transition radius.

If photoevaporation is the dominant mechanism driving the distribution of planet sizes at short orbital periods, then we might expect that closely-spaced planets within multi-planet systems which experience similar irradiation histories would have similar sizes. Kepler-36 is one example to the contrary with both a sub-Neptune and super-Earth orbiting the same star at very similar orbital distances \citep{Carter12}. A detailed analysis of the statistical properties of multi-planet systems utilizing the CKS stellar parameters is currently ongoing (Weiss et al. (in prepration)).

\subsection{Core Mass Distribution}
The masses of planets smaller than Neptune are dominated by the solid core.   Thus, measuring the distribution of core masses provides a valuable constraint on their formation histories. The precise location and depth of the photo-evaporation valley likely depends on the underlying core mass distribution.
Planet masses can be constrained using TTVs \citep{Holman05,Agol05}, but only in specific architectures that may probe different underlying populations. Most of the \Kepler systems studied in this work are faint and out of reach of the current generation of RV instruments. And the number of RV mass measurements for small planets is too small to map out the core mass distribution in fine detail \citep{Howard10, Mayor11}. Teasing out this distrubtion will require a large sample of low-mass planets amenable to mass measurements.  Ongoing and upcoming surveys such as the APF-50 survey \citep{Fulton16}, the HARPS-N rocky planet search \citep{Motalebi15}, MINERVA \citep{Swift15}, and TESS \citep{Ricker14} are working to achieve this goal.

\section{Conclusion}
\label{sec:conclusion}
Using precise planet radii for \val{ncand-cks} \Kepler planets from the CKS Survey, we examined the planet radius distribution at high-resolution. We find evidence for a bimodal distribution of small planet sizes. Sub-Neptunes and super-Earths appear to be two distinct planet classes. Planets tend to prefer radii of either $\sim$1.3 \rearth or $\sim$2.4 \rearthe, with relatively few planets having radii of 1.5--2.0 \rearthe.  Planets in the gap have the maximum size for a rocky core, as seen in previous studies of bulk planet density and of ultra-short period planets.  
We posit that the bimodal planet radius distribution stems from differences in the envelope masses of small planets.
While our current dataset is insufficient to distinguish between theoretical models that produce the gap, it charts a path forward to unraveling further details of the properties of the galaxy's most abundant planets.

\facility{Keck:I (HIRES), Kepler}

\acknowledgments{
The CKS project was conceived, planned, and initiated by AWH, GWM, JAJ, HTI, and TDM. 
AWH, GWM, JAJ acquired Keck telescope time to conduct the magnitude-limited survey. 
Keck time for the other stellar samples was acquired by JNW, LAR, and GWM.
The observations were coordinated by HTI and AWH and carried out by AWH, HTI, GWM, JAJ, TDM, BJF, LMW, EAP, ES, and LAH. 
AWH secured CKS project funding. 
SpecMatch was developed and run by EAP and SME@XSEDE was developed and run by LH and PAC. 
Downstream data products were developed by EAP, HTI, and BJF.
Results from the two pipelines were consolidated and the integrity of the parameters were 
verified by AWH, HTI, EAP, GWM, with assistance from BJF, LMW, ES, LAH, and IJMC.
EAP computed derived planetary and stellar properties with assistance from BJF.  
BJF performed the analysis in this paper, with assistance from EAP, AWH, and GWM.
This manuscript was largely written by BJF with assistance from EAP, AWH, GWM, JNW, and LMW.

We thank Josh Winn, Jason Rowe, Eric Lopez, Jeff Valenti, Daniel Huber, and Leslie Rogers for contributing insight during many helpful conversations and providing comments on early drafts of the manuscript.
Most of the data presented here were determined directly from observations at the W.\ M.\ Keck Observatory, which is operated as a scientific partnership among the California Institute of Technology, the University of California, and NASA. We are grateful to the time assignment committees of the University of Hawaii, the University of California, the California Institute of Technology, and NASA for their generous allocations of observing time that enabled this large project.
Kepler was competitively selected as the tenth NASA Discovery mission. Funding for this mission is provided by the NASA Science Mission Directorate.  
BJF acknowledges that this material is based upon work supported by the National Science Foundation Graduate Research
Fellowship under Grant No. 2014184874. Any opinion, findings, and conclusions or recommendations
expressed in this material are those of the authors and do not necessarily reflect the views of the
National Science Foundation.
EAP acknowledges support from Hubble Fellowship grant HST-HF2-51365.001-A awarded by the Space Telescope Science Institute, which is operated by the Association of Universities for Research in Astronomy, Inc. for NASA under contract NAS 5-26555. 
AWH acknowledges NASA grant NNX12AJ23G.  
TDM acknowledges NASA grant NNX14AE11G.
PAC acknowledges National Science Foundation grant AST-1109612.
LH acknowledges National Science Foundation grant AST-1009810.
LMW acknowledges support from Gloria and Ken Levy and from the Trottier Family.
ES is supported by a post-graduate scholarship from the Natural Sciences and Engineering Research Council of Canada.
Finally, the authors wish to recognize and acknowledge the very significant cultural role and reverence that the summit of Maunakea has always had within the indigenous Hawaiian community.  We are most fortunate to have the opportunity to conduct observations from this mountain.
}

\newpage
\bibliographystyle{aasjournal}
\bibliography{references}

\newpage

\appendix
\renewcommand\thefigure{\thesection.\arabic{figure}}    
\renewcommand\thetable{\thesection.\arabic{table}}    
\setcounter{figure}{0} 
\setcounter{table}{0} 

\section{Non-cummulative Filters}
\label{sec:ap_filters}

\begin{figure*}
\centering
\includegraphics[scale=0.35]{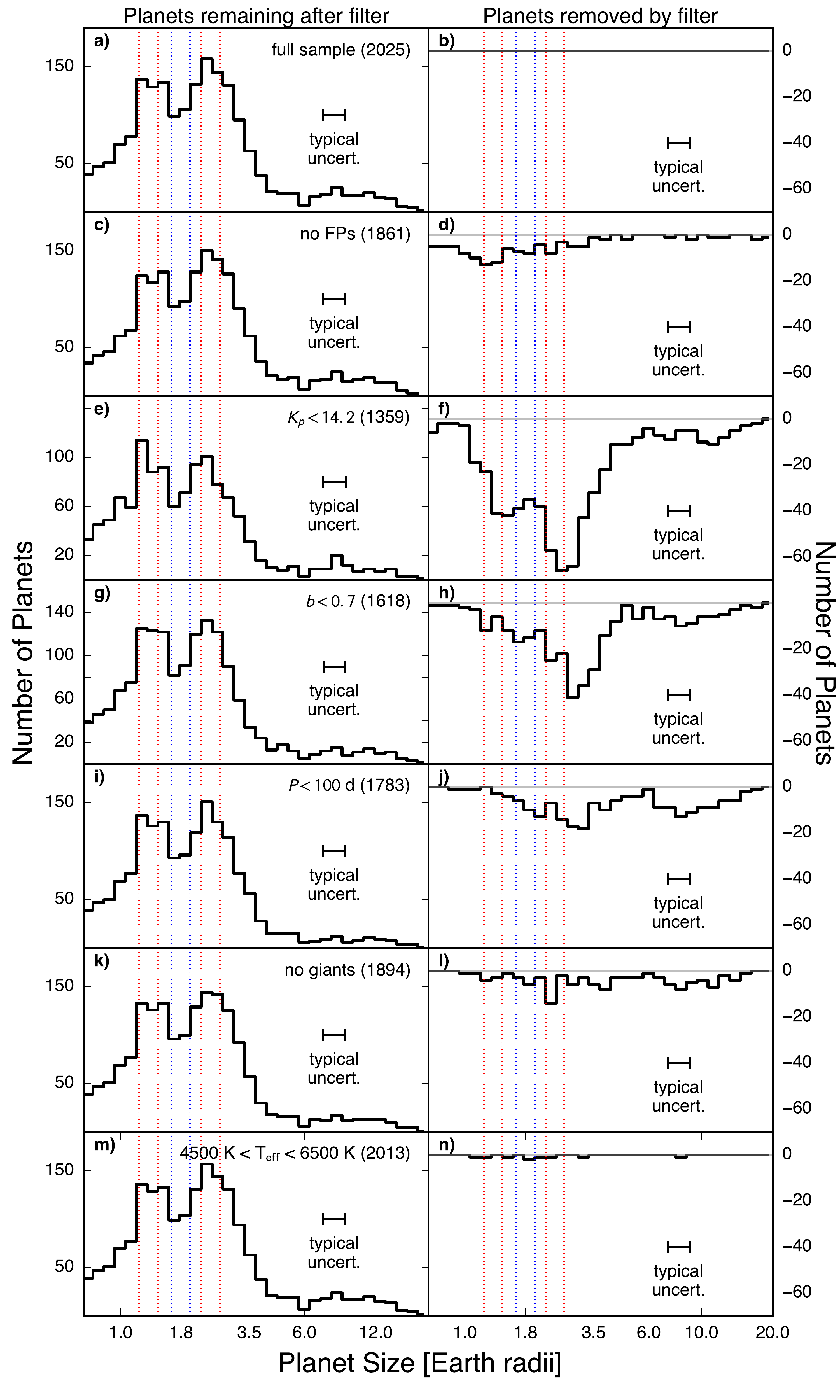}
\caption{{\bf (a)} size distribution of planet candidates from the CKS sample. {\bf (b)} planets removed by the specified filter. Panels {\bf (c)--(n)} show the radius distribution and planets removed from the full sample after applying only a single cut removing known false positives {\bf (c)}, planets orbiting faint host stars {\bf (e)}, planets with grazing transits {\bf (g)}, planets with orbital periods longer than 100 days {\bf (i)}, planets orbiting giant host stars {\bf (k)}, and planets orbiting host stars cooler than 4700 K or hotter than 6500 K {\bf (m)}.
No completeness corrections have been applied. The $b<0.7$ cut and the \Kp $<$ 14.2 cut remove the most planet candidates, but no filter preferentially removes planets in the gap (between blue dotted lines).
}
\label{fig:filters_nc}
\end{figure*}

We investigate the impact of each individual filter on the planet catalog by producing a figure similar to Figure \ref{fig:filters}. However, instead of plotting the distribution after all successive filters are applied to the original sample we plot the distributions after applying \emph{only} the filter specified in the annotations and the figure caption (Figure \ref{fig:filters_nc}). The magnitude and impact parameter cuts have the greatest impact on the final sample since they subtract the greatest number of planets. However, no filter preferentially removes planets in the gap or preferentially preserves planets just outside the gap.

\section{Weighted Kernel Density Estimation}
\label{sec:wKDE}

The weights calculated in Section \ref{sec:completeness} can be used to estimate the occurrence rate distribution of any planet property using weighted kernel density estimation as an alternative to binned histograms \citep[wKDE,][]{Morton14}. We calculate the kernel density estimate as:
\begin{equation}
\phi(x) = \frac{1}{N_{\star}} \sum_{i=1}^{n_{\rm pl}} w_i  \cdot K(x-x_i, \sigma_{x,i}).
\label{eqn:wKDE}
\end{equation}
$K$ is the ``kernel'' and, in general, it can be any non-negative function that integrates to one and has a centroid of zero. $x_i$ are the individual measurements for a given planet property and $\sigma_{x,i}$ are the uncertainties on those measurements. We treat double-sided uncertainties as symmetric Gaussian uncertainties by taking the mean of the reported upper and lower 1-sigma uncertainties. We adopt a standard Gaussian kernel to calculate the one-dimensional distributions of planet properties, and a bivariate Gaussian for two-dimensional distributions. In order to ensure smooth distributions and contours we limit fractional measurement uncertainty to $\geq$5\% in the calculation of the 2D wKDEs. Orbital period is the only parameter that is subject to this limit.

\begin{figure}
\centering
\includegraphics[scale=0.33]{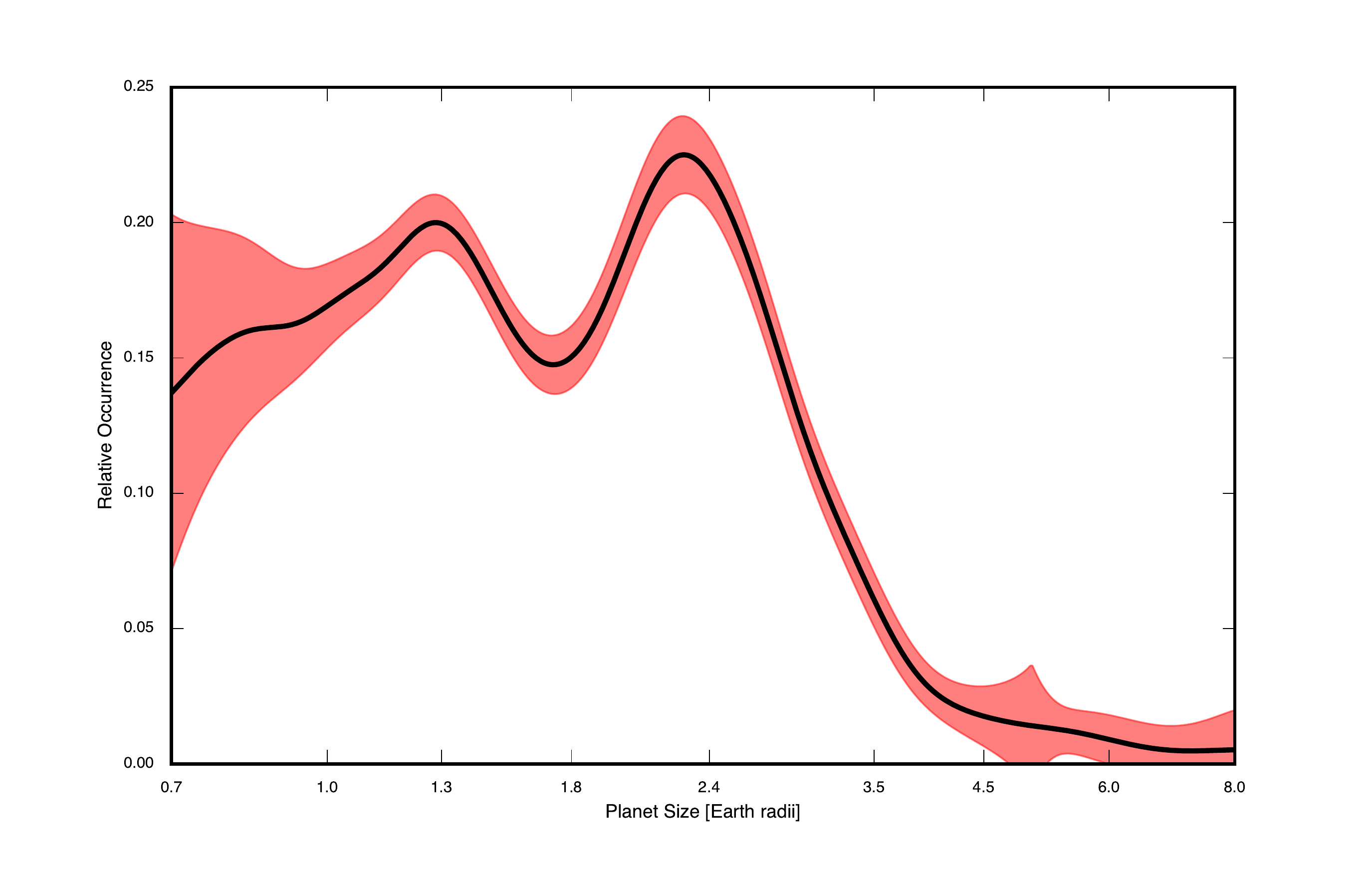}
\caption{Bin-free view of the planet radius distribution calculated using wKDE (Equation \ref{eqn:wKDE}). The 1-sigma uncertainty region is shaded in red and calculated using a suite of simulated transit surveys as described in Appendix \ref{sec:wKDE}. }
\label{fig:kde}
\end{figure}

To investigate the possibility that the gap in the planet radius distribution is an artifact of binning we calculate the planet radius distribution using wKDE (Figure \ref{fig:kde}). We choose a Gaussian kernel and a variable bandwidth that matches the radius uncertainty for each individual measurement. Again, there are two peaks in the radius distribution separated by a gap. The wKDE demonstrates that the presence or location of the gap does not depend on the particular choice of bin size. The contrast between the bottom of the gap and the top of the peaks is reduced in the wKDE-derived planet radius distribution. However, as shown in the simulations described in Appendix \ref{sec:validation}, this is an artifact of the wKDE technique and probably not a good representation of the underlying radius distribution. The planet radius uncertainties are effectively being counted twice in both the scatter of the median values and the width of the Gaussians summed to create the wKDE. The simulations described in Appendix \ref{sec:validation} show the same dilution of the gap depth when using the wKDE to recover known distributions of simulated planets. Quantifying the valley depth from the wKDE radius distribution may require a careful exploration and justification of the kernel bandwidth selection. Our simulations show that the histograms better reproduce the known input distributions, so we choose leave this bandwith tuning for future studies and conclude that the histogram gives a more accurate picture of the planet radius distribution over this particular application and implementation of the wKDE.

\section{Validation of the Completeness Corrections}
\label{sec:validation}
\begin{figure}
\centering
\includegraphics[scale=0.5]{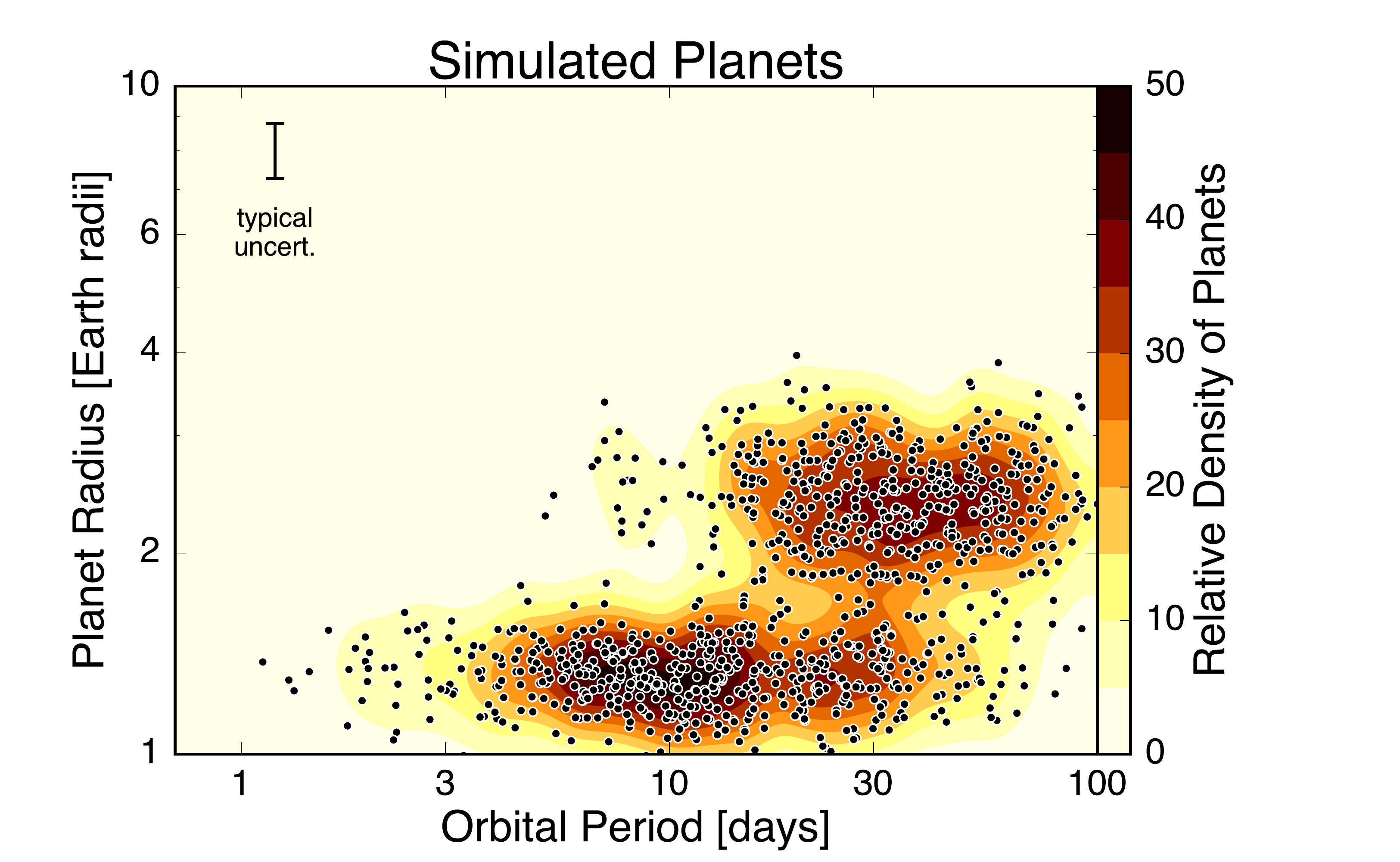}
\caption{Radius vs. period distribution for simulated sample of planets. For plotting clarity and speed we plot only 1,000 randomly chosen simulated planets out of the 45,000 simulated planets.}
\label{fig:sim_planets}
\end{figure}

We validate our occurrence calculations and estimate uncertainties by constructing a suite of 100 simulated transit surveys. For each simulation, we draw a distribution of \val{num-sim-planets} planet radii and orbital periods from two lognormal distributions then sum those distributions together to create a bimodal distribution similar to the distribution observed in our real planet detections (Figure \ref{fig:sim_planets}). We assign each simulated planet to a star in our filtered sample of KOI hosts and calculate detection probabilities and weights as described in \S \ref{sec:occurrence}. These detection probabilities are used to decide which planets would have been detected in our real survey. The number of simulated planets (\val{num-sim-planets}) was chosen such that the mean number of planets in the 100 simulated planet detection catalogues is equal to the total number of planets in our filtered KOI catalogue (\val{num-planets-filtered}).

The stellar radii for the stars in the Stellar17 sample, which are used in the completeness corrections, are perturbed in two different ways in each simulation. We multiply all of the Stellar17 stellar radii by a common constant drawn from a normal distribution centered at 1.0 with a width of 0.25 to simulate potential systematic offsets between the stellar radii in the Stellar17 catalogue and the stellar radii in the CKS catalog. We also add Gaussian noise to the stellar radii for all stars with distribution widths determined from their individual measurement uncertainties. The uncertainties in our final bin heights and occurrence ratios estimated from these simulations account for both systematic and Gaussian random errors in the stellar parameters in the Stellar17 catalog.

\begin{figure}
\centering
\includegraphics[scale=0.5]{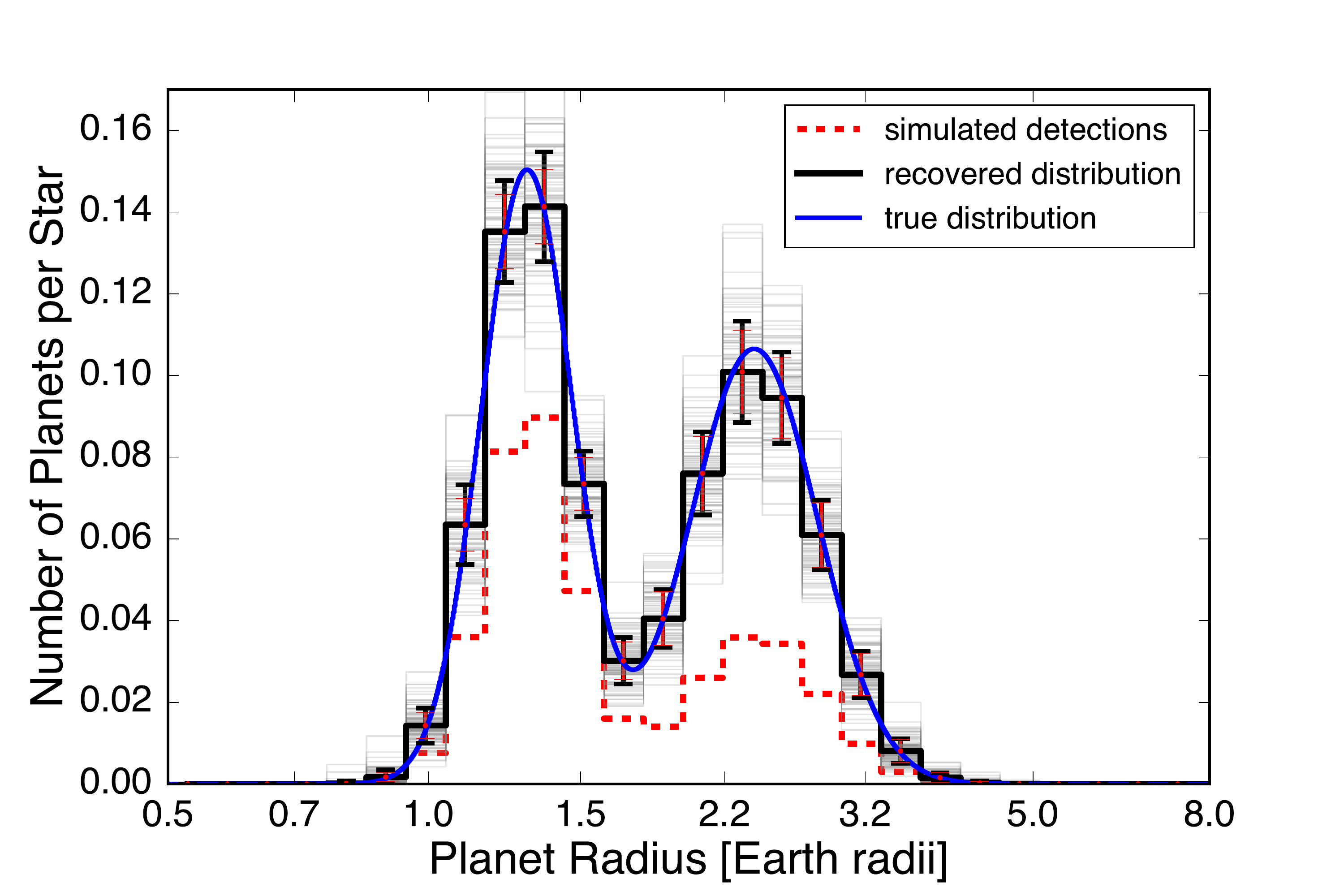}
\includegraphics[scale=0.5]{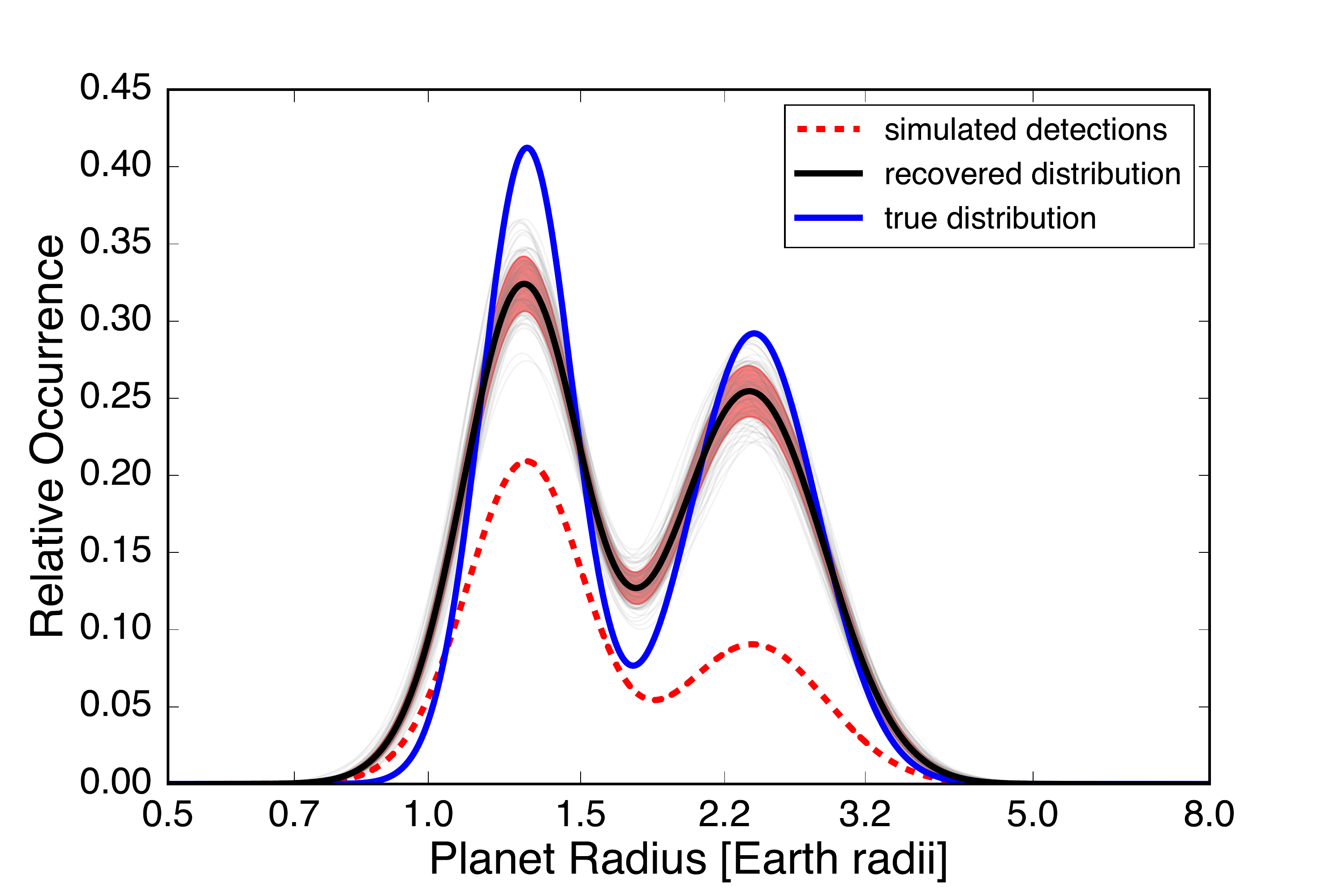}
\caption{\emph{Top:} Results from simulating 100 transit surveys with a known input distribution of planets. The input distribution of simulated planets is plotted in blue,  and the simulated detections are plotted in a red dashed line. The completeness-corrected distributions measured from each of the simulations are plotted as thin grey lines and the median of those recovered distributions is plotted in a thick black line. The thick black error bars are the standard deviation of all of the simulations in each bin and the thin red error bars are poisson uncertainties on the number of detections in each bin scaled by the completeness correction for that bin.
\emph{Bottom:} Same as \emph{top} panel but calculated using the wKDE technique described in \S \ref{sec:occurrence}. The shaded red area encompasses the standard deviation of the resulting wKDEs over all 100 simulations. We adopt this fractional uncertainty for the one-dimensional KDE plotted in Figure \ref{fig:kde}.
}
\label{fig:sim_hist}
\end{figure}

We produce histograms for each simulation and correct them for completeness as described in \S \ref{sec:occurrence}. The standard deviation of the values in each histogram bin become the uncertainty on the bin values. When compared with uncertainties calculated using Poisson statistics on the number of simulated detections in each bin we find that the Poisson uncertainties are underestimated by a factor of \val{error-bin-fudge} depending on the radius bin. In order to avoid small number statistics for the histogram bins where the simulated distribution approaches zero we repeat the simulations with an input distribution of planets that is log-uniform in radius from 0.5--20.0 \rearth and log-uniform in period from 1--200 days solely for the purpose of calculating the uncertainty scaling factors for each radius bin. We adopt the scaling factors listed in Table \ref{tab:unc} in the calculation of all completeness-corrected planet radius histograms and for fitting the distribution described in \S \ref{sec:valley}. 

\begin{deluxetable}{lr}
\tabletypesize{\tiny}
\tablecaption{Bin Uncertainty Scaling Factors}
\tablehead{ 
    \colhead{Radius bin}   &   \colhead{Scaling Factor}  \\
    \colhead{\rearth}        &    \colhead{} } 
\startdata
0.50--0.56  &  2.82 \\
0.56--0.62  &  2.50 \\
0.62--0.69  &  2.30 \\
0.69--0.76  &  2.54 \\
0.76--0.85  &  2.35 \\
0.85--0.94  &  2.09 \\
0.94--1.05  &  1.92 \\
1.05--1.16  &  1.95 \\
1.16--1.29  &  1.89 \\
1.29--1.43  &  1.46 \\
1.43--1.59  &  1.65 \\
1.59--1.77  &  1.81 \\
1.77--1.97  &  1.38 \\
1.97--2.19  &  1.50 \\
2.19--2.43  &  1.39 \\
2.43--2.70  &  1.58 \\
2.70--3.00  &  1.48 \\
3.00--3.33  &  1.58 \\
3.33--3.70  &  1.25 \\
3.70--4.12  &  1.48 \\
4.12--4.57  &  1.47 \\
4.57--5.08  &  1.46 \\
5.08--5.65  &  1.63 \\
5.65--6.27  &  1.45 \\
6.27--6.97  &  1.50 \\
6.97--7.75  &  1.52 \\
7.75--8.61  &  1.34 \\
8.61--9.56  &  1.44 \\
9.56--10.63  &  1.46 \\
10.63--11.81  &  1.52 \\
11.81--13.12  &  1.57 \\
13.12--14.58  &  1.36 \\
14.58--16.20  &  1.35 \\
16.20--18.00  &  1.45 \\
18.00--20.00  &  1.44
\enddata
\label{tab:unc}
\end{deluxetable}

We calculate the occurrence ratio of super-Earths to sub-Neptunes in the same way as we do for the real planet catalogue in \S \ref{sec:rel_occ}. The mean occurrence ratio is consistent with the same ratio for the input distribution of simulated planets and the standard deviation as a fraction of the ratio is 33\%. We adopt this fractional uncertainty for the occurrence ratio calculation on the real planet catalogue.

We also calculate the radius distribution for each simulation using the wKDE technique described in Appendix \ref{sec:wKDE}. We find that the wKDE slightly underestimates the contrast between the peaks of the radius distribution and the bottom of the gap. This is likely due to the fact that there is scatter in the radii measurements due to uncertainties. Those uncertainties are also being included as the widths of the Gaussians used to calculate the wKDE, in effect counting the uncertainty twice. Since we do not perform any quantitative analysis on the wKDE we choose not to ``de-bias'' the wKDE as described in \citet{Morton14}, but instead limit our quantitative analysis to the histograms that seem to be a more accurate representation of the underlying distributions in our simulations. We use the resulting wKDEs from the simulated surveys in order to estimate the fractional uncertainty as a function of planet radius for the wKDE calculated from the real planet catalog (Figure \ref{fig:kde}).

\end{document}